\documentclass[final,3p,times,11pt]{elsarticle}

\pdfoutput = 1
\usepackage{bm}
\usepackage{color}
\usepackage{graphicx}
\usepackage{amssymb,amsmath,amsfonts,amsbsy}
\usepackage{cancel}

{
{ 
{

\def\der{{\rm \ d}}

\DeclareMathAlphabet{\mathpzc}{OT1}{pzc}{m}{it}

\journal{Physics of the Dark Universe Special Issues}

\begin{document}

\begin{frontmatter}

\title{Bayesian analysis of multiple direct detection experiments}

\author{Chiara Arina}
\address{GRAPPA Institute, University of Amsterdam, Science Park 904, 1090 GL Amsterdam (Netherlands)\\ and Institut d'Astrophysique de Paris, Universit\'e Pierre et Marie Curie, 98bis boulevard Arago, 75014 Paris (France)}

\begin{abstract}
Bayesian methods offer a coherent and efficient framework for implementing uncertainties into induction problems. In this article, we review how this approach applies to the analysis of dark matter direct detection experiments. In particular we discuss the exclusion limit of XENON100 and the debated hints of detection under the hypothesis of a WIMP signal. Within parameter inference, marginalizing consistently over uncertainties to extract robust posterior probability distributions, we find that the claimed tension between XENON100 and the other experiments can be partially alleviated in isospin violating scenario, while elastic scattering model appears to be compatible with the frequentist statistical approach. We then move to model comparison, for which Bayesian methods are particularly well suited. Firstly, we investigate the annual modulation seen in CoGeNT data, finding that there is weak evidence for a modulation. Modulation models due to other physics compare unfavorably with the WIMP models, paying the price for their excessive complexity. Secondly, we confront several coherent scattering models to determine the current best physical scenario compatible with the experimental hints. We find that exothermic and inelastic dark matter are moderatly disfavored against the elastic scenario, while the isospin violating model has a similar evidence. Lastly the Bayes' factor gives inconclusive evidence for an incompatibility between the data sets of XENON100 and the hints of detection. The same question assessed with goodness of fit would indicate a $2\sigma$ discrepancy. This suggests that more data are therefore needed to settle this question. 
\end{abstract}


\end{frontmatter}

\section{Introduction}
\label{sec:intro} 

The presence of dark matter, postulated first by Zwicky in 1933 observing the Coma Cluster~\cite{zwicky}, has been nowadays confirmed by several observations in cosmology and astrophysics. Besides precision measurements on its abundance, we only have gravitational evidence for this dark component and its nature and properties are completely unknown. Baryons can constitute only the 4\% of the total content of the Universe, not enough to explain the dark matter content of the Universe ($\sim 30\%$)~\cite{Ade:2013zuv}. This fact points towards a non-baryonic origin for the dark matter and underlines the need for physics beyond the standard model, as neutrinos were relativistic in the early Universe. Several theoretically motivated extensions of the standard model provide dark matter candidates which fall into the category of WIMPs (the most known being the supersymmetric neutralino). As the name indicates, these particles are weakly interacting, massive, neutral and stable at least on cosmological scale, to allow structures we observe to form.  From now on we focus on WIMP cold dark matter, even though other possibilities exist, see {\it e.g.}~\cite{Bertone:2004pz,Bergstrom:2012fi}.

\begin{figure}[t!]
\centering
\includegraphics[width=0.6\columnwidth,trim=5mm 15mm 20mm 10mm, clip]{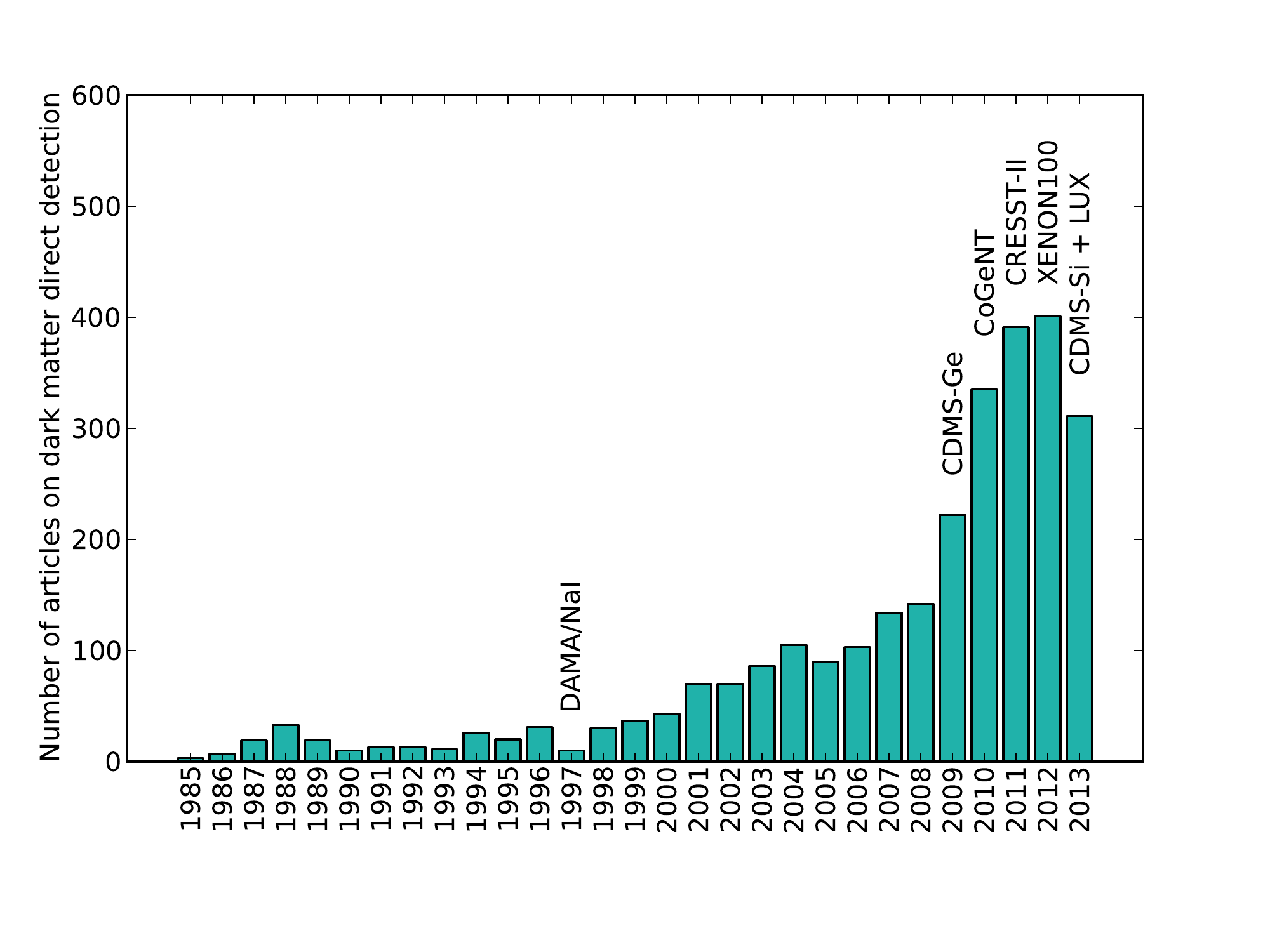}
\caption{{\it The evolution of dark matter direct detection}. Number of articles on this topic as a function of publication year (which is  expected to be larger than the one reported here, source High-Energy Physics Literature Database). The text aligned to a particular year corresponds to a major breakthrough in the field as labelled: the discovery of modulation signal in DAMA/NaI in 1997, the claim of an excess in the CDMS-Ge data in 2009, the claim of a dark matter hint by CoGeNT in 2010, the excess measured by CRESST in 2011, the world best exclusion limit by XENON100 in 2012, the hint for detection by CDMS-Si and the most recent exclusion bound by LUX, both in 2013. Articles until the 1990 focus in particular on design and engineering of the underground detectors. In the nineties literature concentrates in theoretical predictions for specific WIMP candidates ({\it e.g.} neutralino, Kaluza-Klein) and first experimental exclusion bounds become available. After 2000, there is an increasing number of experimental results and consequently a large theoretical effort for reconciling the DAMA excess with upper limits, trend confirmed by the exponential rise in publications in the recent years and concerning the most recent experimental results. On the same vein discussions on astrophysical/experimental uncertainties develop.}
\label{fig:evolDD}
\end{figure} 
There are three basic ways to detect WIMPs: indirect detection, which aims to observe dark matter annihilation products ({\it i.e.} neutrinos, anti-matter and gamma-rays) in the halo regions where the WIMP density is higher. Another possibility is to produce WIMP particles directly at the Large Hadron Collider (LHC) via proton collisions; lastly there is direct detection, which is the central topic of this review. The idea that WIMPs might scatter in an underground detector off nuclei, providing them with detectable recoiling energies, dates back the eighties~\cite{Goodman:1984dc,Drukier:1986tm}. Since then, a huge experimental effort has been deployed, with more than 20 different experiments currently running and reaching unprecedented sensitivities. Several orders of magnitude in the WIMP-nucleus elastic interaction have been constrained by past and current experiments. 

Among all the experiments the most notable exclusion limits  are from XENON100~\cite{Aprile:2012nq} and LUX~\cite{Akerib:2013tjd}, which is currently the strongest bound. Other exclusion limits are provided by EDELWEISS-II~\cite{Armengaud:2012pfa}, ZEPLIN-III~\cite{Akimov:2011tj}, KIMS~\cite{Kim:2012rz}, CRESST commissioning run on W~\cite{Brown:2011dp} and CDMS~\cite{Ahmed:2009zw}. At low-mass WIMP also relevant are the bounds of PICASSO~\cite{Archambault:2012pm}, SIMPLE~\cite{Felizardo:2011uw}, XENON10~\cite{Angle:2011th} and CDMS~\cite{Ahmed:2010wy} low energy analyses. Even though we will not consider further spin-dependent interaction in this review,  we briefly resume the experimental situation as far as it concerns the exclusion limits. The spin-dependent scattering occurs when the dark matter interacts with the spin of the unpaired nucleon (proton or neutron) of the nucleus. COUPP~\cite{Behnke:2012ys} is the most constraining bound for spin-dependent scattering on proton, however other significant exclusion limits come from PICASSO~\cite{Archambault:2012pm}, SIMPLE~\cite{Felizardo:2011uw}, KIMS~\cite{Kim:2012rz}, ZEPLIN-III~\cite{Akimov:2011tj}, CDMS~\cite{Aalseth:2010vx}, XENON10~\cite{Angle:2008we} and XENON100. The neutrino telescope IceCube has as well set stringent limits comparable to those set by COUPP on such interaction by the non observation of neutrinos from the Sun~\cite{Aartsen:2012kia}. Only a few experiments are sensitive to the spin-dependent interaction on neutron: the strongest exclusion bound comes from XENON100~\cite{Aprile:2013doa}, followed by XENON10, ZEPLIN-III and CDMS. 

Alongside the upper bounds, four experiments claim a hint of detection of a light WIMP with a mass in the ballpark $7-12$ GeV: DAMA/LIBRA~\cite{Bernabei:1998fta,Bernabei:2013xsa} (DAMA hereafter), CoGeNT~\cite{Aalseth:2011wp} (both observe as well an annual modulation in the rate, which is supposed to be a smoking gun for dark matter), CRESST~\cite{Angloher:2011uu} and CDMS-II on silicon~\cite{Agnese:2013rvf} (CDMS-Si from now on). These claims are in strong tension with the exclusion bounds if interpreted as due to dark matter scattering off nuclei. In the recent years, the interest in WIMP direct searches increased exponentially, as shown in Fig.~\ref{fig:evolDD}, because of the growing number of experimental results. Theoretical efforts are deployed towards making predictions and discussing the compatibility between detections and exclusion limits, {\it e.g.}~\cite{Belli:1999nz,Savage:2008er,Chang:2010en,Farina:2011pw,Kelso:2011gd,Fornengo:2011sz,Fitzpatrick:2010em,Kopp:2011yr,Frandsen:2011ts,DelNobile:2012tx,Foot:2013msa,Cline:2012ei,Buckley:2013jwa,Feng:2013vod,Hooper:2013cwa,Cotta:2013jna}, while from an experimental point of view, a lot of effort is focused in ameliorating the sensitivity and in trying to confirm or disprove these claims, {\it e.g.}~\cite{Ahmed:2009zw,Ahmed:2010wy,Angle:2011th,Collar:2011kf,Brown:2011dp,Cherwinka:2011ij,Ahmed:2012vq,Amare:2012ex}.

Besides the DAMA evidence for dark matter at $9.3\sigma$ CL (confidence level) after 14 years of running, all the other effects oscillate between $2\sigma$ to $4\sigma$ in significance. These values lie in the `discovery range' of a potential new physical effect and a careful application of statistics can make the difference between claiming a discovery or fitting statistical fluctuations of the background. Indeed it is crucial to properly address a new effect and not rely on the argument that in any case new data will in the future resolve the problem, hence it is not important to bother with a refined statistical analysis. Among a variety of reasons to support this opinion, we stress that it might be that there will be no future decisive data. It is indeed  important to account for uncertainties that affect the measurements, because the complexity of the theoretical models and observations will always increase (besides avoiding theoreticians to explain an effect that is not there in the first place). The uncertainties that affect WIMP direct detection are of two kinds: astrophysical ones, as the properties of the dark matter halo and velocity distribution in the solar neighborhood are poorly known, and experimental ones, related to background discrimination and detector response close to threshold. There has been several attempts to deal with these uncertainties. Regarding the astrophysical ones, two approaches are common. The first method takes them into account by letting the astrophysical parameters vary~\cite{Belli:1999nz,Ullio:2000bf,Green:2002ht,Green:2003yh,Vergados:2007nc,MarchRussell:2008dy,Kuhlen:2009vh,Ling:2009eh,McCabe:2010zh,Schneider:2010jr,Lisanti:2010qx,Belli:2011kw,Fairbairn:2012zs} however uses classical statistical tools to assess their impact. The second approach integrates them out~\cite{Fox:2010bz,HerreroGarcia:2012fu,Gondolo:2012rs,Bozorgnia:2013hsa,Frandsen:2013cna} by making a mapping between the observed signal in one experiment directly into a predicted rate for another one. It becomes possible to directly compare rates in two detectors with different nuclei, however these quantities depend on a third parameter, which is the minimal velocity required to produce a recoil of detectable energy. The experimental systematics are far less investigated, we mention here the few attempts to address the uncertainty on the scintillation efficiency of the XENON100 detector~\cite{Andreas:2010dz,Savage:2010tg,Fornengo:2011sz}.

Bayesian statistics is based on the definition of probability as {\it degree of belief} and on the Bayes' theorem, which is the primary tool for assigning probabilities combining {\it a priori} knowledge and experimental informations. This notion of probability can be applied to any event regardless to the notion of repeated experiments. Bayesian methodology provides at the same time an accurate statistical analysis of the data and addresses uncertainties in the most natural way, as it gives a consistent framework for including them independently of their nature ({\it e.g.} no need of distinguishing between `statistical' or `systematic' uncertainties). Some of the parameters that describe uncertainties affect the data but are of no interest to the analysis of the theoretical model, hence are called {\it nuisance parameters}. Bayesian inference technique deals in a very simple way with the nuisances: it infers the joint posterior probability distribution and then integrates them out, procedure which is called marginalization. In this review we use Bayesian statistics to discuss the status of experimental results in direct dark matter searches, as this approach has not received a large attention from the dark matter direct detection community. Bayesian methods are used in a few papers for forecasting model expectations and WIMP parameter dependence on the astrophysics of the halo~\cite{Strigari:2009zb,Akrami:2010dn,Pato:2010zk,Friedland:2012fa}, while the first application to experimental data can be found in~\cite{Arina:2011si}, followed by~\cite{Arina:2011zh,Arina:2012dr}: these three papers are laying the basis of this review and we refer to them for technical details. Only very recently a Bayesian analysis of XENON100 based on information theory has been proposed by~\cite{Davis:2012hn}, which is alternative to~\cite{Arina:2011si}.  As it will be shown extensively in the paper, the fundamental quantity for Bayesian analysis is the likelihood function for the signal and for the background. The data provided by the LUX collaboration are far from giving enough information to succeed in such task. The collaboration does not provide the spectral information for the background (electronic leakage and neutron recoils), nor a definite quantification of the expected number of events. It is therefore not clear how to treat the events shown in figure 4 of~\cite{Akerib:2013tjd}. Frequentist analyses, which rely on heavy assumptions on the number of seen events, have tried to model the LUX likelihood~\cite{DelNobile:2013gba,DelNobile:2013sia} with partial success, but they all not take into account the background. Hence for the aforementioned reasons, we do not updated our Bayesian analysis with the new LUX data and we demonstrate the Bayesian procedure with the XENON100 data. 

The rest of the review is organized as follows. In Sec.~\ref{sec:bayes} we introduce the key concepts of Bayesian statistics, while in Sec.~\ref{sec:dd} we define the relevant quantities for direct detection rates and WIMP models. Section~\ref{sec:inf} summarizes the status of WIMP parameter inference for a selection of experiments, discussing uncertainties and prior issues. In Sec.~\ref{sec:idc} we apply Bayesian model comparison in two cases: first to assess if CoGeNT data show evidence for a modulated signal due to WIMPs, and second to hunt the best theoretical description of the experiments which have hints of detection; the sensitivity analysis to priors is discussed. Section~\ref{sec:comp} addresses the question of the compatibility of the experimental results. Section~\ref{sec:concl} summarizes our conclusions and gives future perspectives.~\ref{sec:app} contains technical materials on the CDMS-Si data and provide the interested reader with a worked out example of experimental likelihood.

\section{Pills of Bayesian statistics}\label{sec:bayes}
Below we review the main elements of Bayes' theorem, which describes how the state of belief about the model changes after having considered the information provided by the data. For an in-depth discussion see~\cite{Trotta:2008qt} and references therein.

By definition, Bayes' theorem allows to compute the posterior probability distribution (pdf) $\mathcal{P}(\theta | d, {\mathcal M})$ of a given set of parameters $\theta$ defining a model ${\mathcal M}$
\begin{equation}
\label{eq:bt}
\mathcal{P}(\theta | d, {\mathcal M}) = \frac{{\mathcal L}(d|\theta,  {\mathcal M})\,  \pi(\theta|{\mathcal M})}{p(d|{\mathcal M})}.
\end{equation}
Here, $d$ are the data under consideration, ${\mathcal L}(d|\theta,{\mathcal M})$ the likelihood function, and $\pi(\theta|{\mathcal M})$ is the prior pdf for the parameters under the model. The likelihood function encodes the information on how the theoretical model describes the data. The quantity $p(d|{\mathcal M})$, defined as
\begin{equation}
\label{eq:evidence}
{p}(d|\mathcal{M}) \equiv \int \mathcal{L}(d|\theta, \mathcal{M})\, \pi(\theta|\mathcal{M}) \der \theta\,,
\end{equation}
is called the Bayesian evidence.

\subsection{Prior choice}\label{sec:prior}
Bayes' theorem requires to specify $\pi(\theta|\mathcal{M}) $, the probability density on the parameter space $\theta$ prior to the observation of the data $d$. Since this prior pdf is independent of the data, it needs to be chosen according to one's theoretical belief on the model. Often no unique theoretically motivated prior pdf can be derived, hence one may wish to use one which does not favor any parameter region in particular. There are two common choices, depending on the theoretical parameter value range. First, there is the uniform prior
\begin{equation}
\pi_{\rm flat}(\theta|\mathcal{M})  \propto \left\{ 
    \begin{array}{cl} 
     1, &
     {\rm if}\ \theta_{\rm min} \leq \theta \leq  \theta_{\rm max} ,
     \\ 0, & {\rm otherwise},
    \end{array}
\right.
\end{equation}
if the general order of magnitude of the parameter is known. Second, if the order of magnitude is unknown, one may want to use a log-prior instead,
\begin{equation}
\pi_{\rm log}(\theta|\mathcal{M})  = \left\{ 
    \begin{array}{cl} 
   1/\theta, &
      {\rm if}\ \theta_{\rm min} \leq \theta \leq  \theta_{\rm max},
      \\ 0, & {\rm otherwise},
    \end{array}
\right.
\end{equation}
which is equivalent to a uniform prior on $\log \theta$.  Note that because the volume element $ \der \theta$ is in general not invariant under a parameter transformation $f: \theta \to \theta'$, a uniform prior pdf on $\theta$ does not yield the same probabilities as a uniform prior pdf on $\theta'$ unless the mapping $f$ is linear\footnote{The same is also true for the posterior probabilities we discuss in Sec.~\ref{sec:pi}, {\it i.e.} $  \mathcal{P} (\theta | d) {\der}\theta \neq \mathcal{P} (\theta' | d) {\der}\theta'$ in general.}. In both cases the limits $\theta_{\rm min}$ and $\theta_{\rm max}$ should be chosen such that they are well beyond the parameter region of interest.

\subsection{Parameter Inference}\label{sec:pi}

The posterior pdf represents our state of knowledge about the parameters after taking into account the information contained in the data, and its intuitive interpretation is that $\int_V \mathcal{P} (\theta | d) {\rm d}\theta$ gives the probability that the true value of $\theta$ lies in the volume $V$.  While the posterior pdf contains all the necessary information for the interpretation of  the data, one might be interested in its projection into a region of smaller dimensionality and being a function of the relevant theoretical parameters only. By virtue of being a probability
density, its dimensionality can be easily reduced by integrating out the nuisance parameter directions $\psi_i$, yielding an $n$-dimensional marginal posterior pdf,
\begin{equation}
 \label{eq:marg}
 \mathcal{P}_{\rm marg}(\theta_1, ..., \theta_n | d) \propto \int \der\psi_1
 ... \der\psi_m \ {\cal P}( \theta_1, ..., \theta_n,\psi_1...,
 \psi_m|d) \,, 
\end{equation} 
which is more amenable to visual presentation if $n=1,2,3$ and can be used to construct constraints on the remaining parameters.

A complementary approach to marginalization is to project $\mathcal{P}(\theta|d)$ onto the $n$-dimensional subspace by maximizing along the nuisance directions
\begin{equation}
 \label{eq:profile}
 \mathcal{P}_{\rm prof}(\theta_1, ..., \theta_n | d) \propto  \max_{\psi_1
 ... \psi_m} \ {\cal P}( \theta_1, ..., \theta_n,\psi_1...,
 \psi_m|d) \,.
\end{equation}
Maximization is not a Bayesian procedure, and the resulting profile posterior cannot be interpreted as a probability density function. Only for the choice of uniform priors, $\mathcal{P}_{\rm prof}$ coincides with the profile likelihood encountered in classical statistics. However, because $\mathcal{P}_{\rm prof}$ is by construction insensitive to volume effects, it can be used to assess if the inference has been significantly affected by the choice of nuisance parameterization.

\subsection{Bayesian evidence and model comparison}\label{sec:mctheo}
Bayesian inference is based on the assumption that the model $\mathcal{M}$ under consideration is the correct one. However one might want to know what is the viability of the model itself, or rather, among a set of possible alternative models, which one performs better in explaining the data $d$: this is the subject of Bayesian model comparison. In a classical sense such comparison is not possible, as there is no notion of `ranking' models but of hypotheses rejection only. Conversely within Bayesian statistics, model comparison comes out naturally as it incorporates the quantitative notion of Occams' razor by means of the evidence, Eq.~(\ref{eq:evidence}) (which is the average  of the likelihood under the prior for a specific model). Models with excessive complexity, unsupported by the data, are penalized for wasted parameter space. Increasing the dimensionality of the parameter space without significantly enhancing the likelihood $\mathcal{L}(d|\theta, \mathcal{M})$ in the new parameter directions reduces the evidence. Unpredictive priors $\pi(\theta|\mathcal{M})$ ({\it e.g.} excessively broad compared with the width of the likelihood) likewise dilute the evidence.

The posterior probability $\mathcal{P}(\mathcal{M}|d)$ of a model $\mathcal{M}$ is related to the Bayesian evidence via Bayes' theorem,
\begin{equation}
\mathcal{P}(\mathcal{M}|d) \propto p(d|\mathcal{M})\  \pi(\mathcal{M}) \,,
\end{equation}
where $\pi(\mathcal{M})$ is the prior probability assigned to the model $\mathcal{M}$ itself, and we have dropped a normalization constant corresponding to the probability of the data $d$. Namely we can compute the posterior probability of the model, given the experimental data, providing accordingly an update of our belief in each of the theoretical models in the light of the observations.

The posterior odds between two competing models $\mathcal{M}_0$ and $\mathcal{M}_1$ are given by
\begin{equation}
\frac{\mathcal{P}(\mathcal{M}_1|d)}{\mathcal{P}(\mathcal{M}_0|d)} = B_{10}\frac{\pi(\mathcal{M}_1) }{\pi(\mathcal{M}_0) }\,,
\end{equation}
where  
\begin{equation}
\label{eq:bayesfactor}
B_{10} \equiv \frac{p(d|\mathcal{M}_1)}{p(d|\mathcal{M}_0)}\,
\end{equation}
is the Bayes factor, defined as a ratio of the models' evidences.

The Bayes factor represents an update from our prior belief in the odds of two competing models $\pi(\mathcal{M}_1)/\pi(\mathcal{M}_0)$
to the posterior odds $\mathcal{P}(\mathcal{M}_1|d)/\mathcal{P}(\mathcal{M}_0|d)$. If the prior over models is non-committal ({\it i.e.} $\pi(\mathcal{M}_1)=\pi(\mathcal{M}_0)$) the Bayes factor alone determines the outcome of the model comparison.
A Bayes factor larger than unity means that the model $\mathcal{M}_1$ is preferred over the model $\mathcal{M}_0$ as a description of the 
experimental data, and vice-versa. As it may be expected, the use of the Bayes factor as a decision-making criterion is a matter of 
convention\footnote{Similarly to the convention for which the classical statistics criterion rejects a null hypothesis if the p-value falls below
 0.05.} and a common choice is the  Jeffreys' scale, shown in Tab.~\ref{tab:jef}.

\begin{table}[t!]
\caption{Jeffreys' scale for grading the strength of evidence for two competing models $\mathcal{M}_0$ and $\mathcal{M}_1$, here slightly modified  from~\cite{Gordon:2007xm,Trotta:2008qt}.\label{tab:jef}}
\begin{center}
\begin{tabular}{lll}
\hline
$\ln B_{10}$ & Odds $\mathcal{M}_1: \mathcal{M}_0$& Strength of evidence \\
\hline
$<-5.0$ & $< 1:150$ & Strong evidence for $\mathcal{M}_0$ \\
$-5.0 \to -2.5$  & $1:150 \to 1:12$ & Moderate evidence for $\mathcal{M}_0$ \\
$-2.5 \to -1.0$ & $1:12 \to 1:3$ & Weak evidence for $\mathcal{M}_0 $ \\
$-1.0 \to 1.0$ & $1:3 \to 3:1$ & Inconclusive\\
$1.0 \to 2.5$ & $3:1 \to 12:1$ & Weak evidence against $\mathcal{M}_0 $ \\
$2.5 \to 5.0$  & $12:1 \to 150:1$ & Moderate evidence against $\mathcal{M}_0$ \\
$> 5.0$ &$> 150:1$ & Strong evidence against $\mathcal{M}_0$ \\
\hline
\end{tabular}
\end{center}
\end{table}

The prior pdf for model comparison should be carefully chosen, as the evidence is sensitive to its volume; because its choice is usually not unique, interpretation of the results of Bayesian model selection ought to allow for the impact of a reasonable change of priors. This is called {\it sensitivity analysis}. If the models $\mathcal{M}_0$ and $\mathcal{M}_1$ are nested and their parameter priors separable, then the impact of changing the prior width on the Bayes factor can be estimated analytically using the Savage-Dickey density ratio (SDDR, see~\cite{Trotta:2005ar}). The rough idea beyond this ratio is that the prior pdf is normalized hence an increase in its width will lead to a decrease in the posterior  pdf and therefore to a smaller Bayes factor.  If for instance the prior pdf is a top-hat function, the SDDR formula shows that rescaling its width by a factor $\alpha$ will change $\ln B_{10}$ by approximately $-\ln \alpha$. Further on, we discuss how to use this analytic approximation to perform a sensitivity analysis of model comparison results.

\subsection{Compatibility of data sets}\label{sec:ct}

The Occams' razor notion encoded in the Bayesian evidence definition can be used to perform a consistency test between two or more data sets (see {\it e.g.}~\cite{Feroz:2008wr}). Outcomes from different experiments may privilege different corners of the parameter space because of their discrepant data: the compatibility test provides an actual quantitative measure of  the disagreement, alternative to the `chi by eye' assessment from classical statistical tools. 

A data set $d$ can be divided into two parts as $d=\{\mathcal{T},\mathcal{D}\}$, where $\mathcal{T}$ is the subset to be tested for compatibility with respect to the remaining data set $\mathcal{D}$, which we believe to be correct. The conditional evidence $p(\mathcal{T}|\mathcal{D})$ is given by the probability of measuring the data $\mathcal{T}$, knowing that the set $\mathcal{D}$ has been measured
\begin{equation}\label{eq:jointpdf}
p(\mathcal{T}|\mathcal{D}) = \frac{p(\mathcal{T},\mathcal{D})}{p(\mathcal{D})}\,.
\end{equation}
Here $p(\mathcal{T},\mathcal{D})$ is the joint evidence, that is the probability of measuring the whole set $d$ within the model under investigation. This measure is independent on the model parameters $\theta$, as they have been integrated out. Then $p(\mathcal{D})$ is the Bayesian evidence corresponding only to the data subset $\mathcal{D}$ and is a normalization factor that will cancel out. Note that the conditioning on the model $\mathcal{M}$ is understood in this section. The two common definitions for the compatibility tests are as follows.

The $\mathcal{R}$-test, called model comparison test, is an extension of the concept of model comparison in data space to investigate hypotheses. Suppose that $\mathcal{H}_0$ states that all the data sets under scrutiny are compatible with each other and with the models' assumption. On the contrary $\mathcal{H}_1$ affirms that the observed experimental outcomes are inconsistent  so that each data set requires its own set of parameter values, as they privilege different regions in the parameter space. Then the Bayes factor between the two hypotheses, if we have no reason to prefer either $\mathcal{H}_0$ or $\mathcal{H}_1$, is given by
\begin{equation}\label{eq:rtest}
\mathcal{R}(\mathcal{T}^{\rm obs}) = \frac{p(\mathcal{T}^{\rm obs}, \mathcal{D} | \mathcal{H}_0)}{p(\mathcal{T}^{\rm obs} | \mathcal{H}_1) p(\mathcal{D} | \mathcal{H}_1)}\,.
\end{equation}
The strength of evidence against/in favor of $\mathcal{H}_0$ is assessed in the same way as for Bayesian model selection by using the Jeffreys' scale (Tab.~\ref{tab:jef}).
 
The predictive likelihood test or $\mathcal{L}-$test has an interpretation similar to classical hypothesis testing and is defined as follows. The consistency between $\mathcal{T}^{\rm obs}$, defined to be the observed value for the variable $\mathcal{T}$, and $\mathcal{D}$, is evaluated by taking the ratio of  $p(\mathcal{T}^{\rm obs}|\mathcal{D})$ and $p(\mathcal{T}^{\rm max}|\mathcal{D})$, where $\mathcal{T}^{\rm max}$ is the value that maximizes such probability
\begin{equation}\label{eq:ltest}
\mathcal{L}(\mathcal{T}^{\rm obs}|\mathcal{D}) = \frac{p(\mathcal{T}^{\rm obs}|\mathcal{D})}{p(\mathcal{T}^{\rm max}|\mathcal{D})} = \frac{p(\mathcal{T}^{\rm obs},\mathcal{D})}{p(\mathcal{T}^{\rm max},\mathcal{D})}\,.
\end{equation}
By using Eq.~(\ref{eq:jointpdf}), the $\mathcal{L}(\mathcal{T}|\mathcal{D})$ distribution is simply given by the ratio of the joint evidences at the observed and maximal value. To construct concretely such a distribution, one has to evaluate the joint evidence as a function of the possible outcome of the observation $\mathcal{T}$ keeping fixed $\mathcal{D}$. In other words, we vary the value of  $\mathcal{T}$ (assuming the same errors on systematics as reported by the experiment) over a range of values and we compute $\mathcal{L}(\mathcal{T}|\mathcal{D})$ at each value. We then take the value that maximizes this distribution to measure the relative probability of obtaining the observed data realization $\mathcal{T}^{\rm obs}$, as given by Eq.~(\ref{eq:ltest}). If the outcome of the comparison, $\ln \mathcal{L}(\mathcal{T}^{\rm obs}|\mathcal{D})$, is close to zero both data sets are compatible with each other and with the model assumptions. If however $\ln \mathcal{L}(\mathcal{T}^{\rm obs}|\mathcal{D}) \ll 0$, there is a tension between $\mathcal{D}$ and $\mathcal{T}^{\rm obs}$. This means that one should doubt the models' assumption or doubt $\mathcal{T}^{\rm obs}$ (or vice-versa doubt  $\mathcal{D}$) and look properly for systematics. The $\mathcal{L}$-test is weakly dependent on the prior choice, being a likelihood ratio by definition in data space (namely integrated over all possible values of the models' parameters) and is evaluated on a similar significance scale as $\Delta \chi^2$.

\subsection{Discussion on numerical tools}\label{sec:num}

From a practical point of view, the sampling of the posterior pdf and hence of the theoretical parameter space requires appropriate numerical techniques, as the process of marginalization involves the evaluation of a multi-dimensional integral. For Bayesian inference only, Markov Chain Monte Carlo (MCMC) methods based on the Metropolis-Hastings algorithm~\cite{Metropolis53,Hastings70} provide an accurate sampling of both the posterior density function and the profile likelihood and are widespread in cosmology (and start to be used in high-energy physics). For instance the analysis in~\cite{Arina:2011si} uses an appropriately modified version of the public MCMC code \texttt{CosmoMC}~\cite{Lewis:2002ah}. Multimodal nested sampling algorithms are adequate for computing the Bayesian evidence and we mention here the publicly available package \texttt{MultiNest}~\cite{Feroz:2007kg,Feroz:2008xx}. This algorithm has two basic advantages with respect to a MCMC: first it computes the evidence of the model, secondly it is more efficient for `problematic' posterior distributions ({\it e.g.} posterior pdfs which are multimodal or that exhibit pronounced degeneracies). For the same number of parameters, \texttt{MultiNest} reduces the computational time to evaluate the posterior distribution with respect to a MCMC. The resulting chains can be analyzed with the \texttt{GetPlot} or \texttt{GetDist} packages in \texttt{SuperBayes}~\cite{Trotta:2008bp,superbayes} and \texttt{CosmoMC} respectively. A further improvement is \texttt{BAMBI}~\cite{Graff:2011gv}, which combines the neural network training algorithm \texttt{SkyNet}~\cite{Graff:2013cla} with \texttt{MultiNest} for a even faster likelihood evaluation, providing an additional speedup in the posterior pdf sampling. 

\section{Dark matter direct detection rate and theoretical models}\label{sec:dd}
The differential spectrum for nuclear recoils arising from the scattering of WIMPs off target nuclei, in units of cpd/kg/keV (counts per day per detector mass per energy), has the form
\begin{equation}
\label{eq:diffrate}
\frac{\der R}{\der E} = \frac{\rho_{\odot}}{m_{\rm DM} M_N}  \int_{v>v_{\rm min}} \der^3v  \, \frac{\der\sigma}{\der E} \, v  \, f (\vec{v}(t))\,,
\end{equation}
where $E$ is the energy transferred during the collision,  $\rho_{\odot} \equiv \rho_{\rm DM}(R_{\odot})$ the WIMP density at the Sun position, $m_{\rm DM}$ the WIMP mass, $M_{\cal N}$ the mass of the target nucleus, $\der \sigma/\der E$ the differential scattering cross section, and $f(\vec{v}(t))$ is the WIMP velocity distribution in the Earth's rest frame, normalized to unity. The integration in the differential rate is performed over all incident particles capable of depositing a recoil energy of $E$, or equivalently having a velocity larger than $v_{\rm min}$. For pure spin-independent (SI) interaction, which we consider in this review, the differential cross-section encodes the particle physics and nuclear model as
\begin{equation}
\label{eq:pppart}
\frac{\der \sigma}{\der E} = \frac{M_{\cal N} \sigma^{\rm SI}_n}{2 \mu^2_n {v}^2}\ \frac{\Big(f_p Z + (A-Z) f_n\Big)^2}{f_n^2} {\cal F}^2( E) \,  ,
\end{equation}
where $\mu_n=m_{\rm DM} m_n/(m_{\rm DM}+m_n)$ is the WIMP-nucleon reduced mass, $\sigma^{\rm SI}_n$ the SI zero-momentum WIMP-nucleon cross-section, $Z$ ($A$) the atomic (mass) number of the target nucleus, and $f_p, f_n$ are the WIMP effective coherent couplings to the proton and neutron respectively. The nuclear form factor ${\cal F}(E)$ characterizes the loss of coherence for nonzero momentum transfer and is parametrized with the Helm form factor for all nuclei~\cite{Helm:1956zz}.  For a good introduction on direct detection we refer to~\cite{Lewin:1995rx}, while for details about SI form factors we refer to~\cite{Duda:2006uk} and references therein. 

By means of the Bayes' theorem, Eq.~(\ref{eq:bt}), the inference and model comparison problems are explicitly stated once we provide the theoretical model and the likelihood function for the experiments under consideration. 

\subsection{Models for coherent WIMP-nucleus scattering}\label{sec:mcs}
Among all the SI interactions that have been proposed to reconcile the debated experimental results, we review the Bayesian statistical analysis of the four most widespread models.
\begin{enumerate} 
\item Elastic scattering 

This is the standard WIMP interaction which assumes equal coupling to proton and neutron\footnote{This is a reasonable assumption verified in most of particle physics models, such as supersymmetry, universal extra dimensions, Higgs portal.}, namely $f_p=f_n$ in Eq.~(\ref{eq:pppart}). Hence it emerges a quadratic dependence on $A$: heavy nuclei will be more sensitive to heavy WIMP and vice-versa. The minimal velocity to produce a recoil of energy $E$ is simply given by kinematics and requires  $v_{\rm min} = \sqrt{M_{\cal N} E/2 \mu^2}$, where $\mu=m_{\rm DM} M_{\cal N}/(m_{\rm DM}+M_{\cal N})$ is the WIMP-nucleus reduced mass. 

\item Inelastic scattering~\cite{TuckerSmith:2001hy}

The WIMP $\chi$ interacts with the nucleus and jumps into a heavier excited state: $\chi \mathcal{N} \to \chi^{\ast} \mathcal{N}$. The scatter occurs only if the splitting in mass (called $\delta$) between the dark matter mass at the ground state and the excited state, is relatively small ($\mathcal{O}({\rm keV})$). The $v_{\rm min}$ gets modified as  
\begin{equation}
v_{\rm min} = \sqrt{\frac{1}{2 M_{\mathcal N} E_R}} \Big(\frac{M_{\mathcal N} E_R}{\mu}+\delta\Big)  \,.
\label{eq:vmininel}
\end{equation}
Only particles in the very high tail of the velocity distribution will have enough energy to produce a recoil in the detector, hence heavy nuclei will be particularly sensitive to this interaction. 

\item Exothermic dark matter~\cite{Graham:2010ca}

This is exactly the same interaction as inelastic scattering with however a negative $\delta$, meaning that the WIMP after the scattering makes a transition to a lower mass state (explaining as well the name of the interaction).

\item Isospin violating scattering~\cite{Feng:2011vu} 

This model relies on the hypothesis that the WIMP-neutron and WIMP-proton interaction is of different strength, namely $f_n \neq f_p$ in Eq.~(\ref{eq:pppart}), while minimal velocity is defined as for the elastic interaction.
\end{enumerate}

As far as it concern other type of interactions, an exhaustive list of coherent scattering operators, using a non relativistic approach, is provided in~\cite{DelNobile:2013sia}. We mention as well WIMPs that have electromagnetic interaction, leading to {\it e.g.} long range~\cite{Fornengo:2011sz}, or anapole and magnetic interactions~\cite{Pospelov:2000bq,Ho:2012bg,Cline:2012is}. In all these cases, the standard formula for the differential rate, Eq.~\ref{eq:diffrate}, has to be slightly modified accordingly to the interaction or generalized to include the appropriate form factors, velocity dependence and/or momentum dependence.

\subsection{Likelihood and Uncertainties}\label{sec:unc}

The likelihood function ${\cal L}(d|\theta,\mathcal{M})$ describes our belief on how the theoretical parameters $\theta$ of a given model $\mathcal{M}$ connect with the experimental data $d$. Here the experimental quantity of interest is the number of nuclear recoils that occurred in a given observed energy range $[{\cal E}_1,{\cal E}_2]$, which we call $N_{\rm obs}$. The total number of recoils expected in a detector from WIMP-nucleus interaction is obtained by integrating Eq.~(\ref{eq:diffrate}) over energy
\begin{equation}
\label{eq:totrate}
S(\theta)  = M_{\rm det} T \int_{{\cal E}_1/q}^{{\cal E}_2/q}  \der E\  \frac{\der R}{\der E} \,,
\end{equation}
where $M_{\rm det}\, T$ denotes the detector mass times the exposure time, $q$ is the quenching factor, described below, and $\theta = m_{\rm DM}, \sigma ^{\rm SI}_n, f_n/f_p, \delta, ...$\, are the theoretical parameters of interest. 

Contrary to what happens in cosmology, the direct detection collaborations do not release the likelihood functions they use for the data analysis. Hence the first step of the analysis is to write down a likelihood function which is as close as possible to the experimental one, with all possible systematics folded in (provided the information given in the experimental papers). We do not detail the likelihood function for each experiment, but we refer to~\cite{Arina:2011si,Arina:2012dr} and to~\ref{sec:app} for the CDMS-Si likelihood function; here we only sketch the guidelines for its construction. 

For all experiments where $N_{\rm obs}$ is small (typically it oscillates between 0 and 4 events) the likelihood function is given by a Poisson distribution, while for DAMA it is fair to assume that the large number of data is Gaussian distributed. This experiment and CoGeNT are in addition sensitive to the annual modulation, which is a smoking gun signature for dark matter proposed in~\cite{Freese:1987wu,Spergel:1987kx}. It relies upon the idea that because of the movement of the Earth around the Sun, the  local dark matter velocity distribution and the rate become time-dependent 
\begin{equation}\label{eq:mod}
R_{\rm m}(t) \propto S_{\rm m} \cos[2 \pi (t-t_0)/T]\,.
\end{equation}
The WIMP modulated rate $R_m(t)$ has a sinusoidal behavior with a period $T$ of one year, and a phase $t_0$ determined by the movement of the Earth with respect to the Galactic frame (for standard isotropic halo $t_0=2^{\rm nd}$ June). The likelihood function of DAMA and CoGeNT takes into account the modulated rate, proportional to $S_{\rm m}$, with a Gaussian distribution.

The systematics depend on the details of the experiment but can mainly be subdivided into the following categories :
\begin{enumerate}
\item[1.] Background

Every experiment has background events. For direct detection the most dangerous backgrounds are neutron recoils or surface/zero charge events from electron recoils, which all can mimic the nuclear recoil produced by a WIMP. Denoting the expected background rate by $B$ and considering it as nuisance parameter,  the total signal in the likelihood to be confronted  with $N_{\rm obs}$ is then $S(\theta)+B$. As the background comes with an uncertainty $\sigma_B$, it is natural to construct a likelihood function for it, which is however a  detector-dependent quantity. When possible, it is useful to marginalize analytically over the background to reduce the dimensionality of the joint posterior pdf.
\item[2.] Quenching factors (concerning scintillators, such as DAMA, CRESST)

The quenching factor $q$, defined via ${\cal E}=q E$, denotes the fraction of recoil energy that is ultimately observed in a specific detection channel. To distinguish ${\cal E}$ from the actual nuclear recoil energy $E$, the former is usually given in units of keVee (electron equivalent keV), while the latter in keVnr (nuclear recoil keV) or simply keV. The measurement of such quantities is delicate and is affected by large uncertainties. One should then consider the quenching factors as nuisance parameters and vary them with flat priors over the whole experimental allowed range.
\item[3.] Scintillation efficiency in XENON100 (and in two phases nobel gas detectors)

This parameter is a conversion factor between photo-electron (PE) measured by photo-multipliers in the detector and the actual nuclear recoil energy released during the collision. As for the quenching factors, this is a tricky measurement, in particular close to the threshold of 3 PE. The exact dependence of $L_{\rm eff}$ below threshold is not well measured~\cite{Aprile:2012an} and affects the exclusion bound because of Poisson fluctuations~\cite{Andreas:2010dz,Savage:2010tg,Fornengo:2011sz}: different parametrizations of ${\rm L_{eff}}$ can either enhance or reduce the compatibility between the exclusion limits and the DAMA/CoGeNT/CRESST/CDMS-Si preferred parameters. The best fit behavior of course lies in between the extreme cases, however the likelihood function should take this range of possibilities into account, for instance with the addition of one nuisance parameter Gaussian distributed~\cite{Aprile:2011hx,Arina:2011si}.
\item[4.] Nuclear uncertainties

We briefly mention here that the nuclear form factors can introduce considerable uncertainties in the total number of events predicted in a detector when spin-dependent interaction is considered~\cite{Cerdeno:2012ix,Arina:2013jya}, however for SI interaction these uncertainties are tiny and are not considered in the analysis.
\end{enumerate}

As different experiments are independent, the log-likelihood for each observation simply add 
\begin{equation}
- \ln \mathcal{L}_{\rm tot} = - \left(\, \ln \mathcal{L}_{\rm XENON100} + \ln\mathcal{L}_{\rm CDMS-Si} + \ln\mathcal{L}_{\rm DAMA} + ...\, \right)\,.
\end{equation}
However care must be taken when looking at the outcome: the posterior pdf can single out a region that is not favored by any of the experiments, which is a quite meaningless result, as the data sets might be mutually incompatible under the theoretical hypothesis. We will discuss this issue in Sec.~\ref{sec:idc}.

The astrophysical uncertainties are common to all experiments and are
\begin{enumerate}
\item[1.] Shape of the velocity distribution of the dark matter in the halo
 
Most of the analyses parametrize $f(v)$, Eq.~(\ref{eq:diffrate}), with a Maxwellian distribution (the so-called standard halo model SHM), but this has no justification a priori other than being the simplest model that can be handled analytically. The approach used in~\cite{Arina:2011si,Arina:2012dr} computes the velocity distribution directly from the dark matter density profile (NFW, Einasto,Burkert and cored isothermal), under the assumption of spherical symmetry and isotropy, by means of the Eddington formula~\cite{Binneybook,Arina:2011si}. An alternative approach is to use a parametric form for $f(v)$ coming from N-body simulations of galaxy size objects~\cite{Lisanti:2010qx,Mao:2013nda}, which has the advantage of incorporating anisotropies seen in the numerical simulations. We will not discuss further this second method.

\item[2.] Astrophysical parameters

The values at the Sun position of these parameters, which are the circular velocity $v_0$, the escape velocity $v_{\rm esc}$ and $\rho_\odot$, are known up to a certain degree of precision, ranging from 20\% to a factor of 2. Instead of keeping them fixed at their preferred values, they should be folded into the analysis as nuisance parameters.
\end{enumerate}

\subsection{Discussion on prior pdfs}\label{sec:dp}

The main parameters of interest are $m_{\rm DM}$ and $\sigma_n^{\rm SI}$: they are common to all particle physics model considered above. These are accompanied by a set of astrophysical and experiment-specific systematic nuisance parameters plus additional theoretical parameters for the more complicated models, inelastic ($\delta > 0$), exothermic ($\delta < 0$) or isospin violating ($f_n/f_p$) dark matter. We discuss here the prior choice for the parameters following~\cite{Arina:2011si,Arina:2012dr}, where the details can be found.

In specifying prior pdfs for $m_{\rm DM}$ and $\sigma_n^{\rm SI}$, we can only rely on the assumption that the dark matter particle is a WIMP. In other words the mass and the interaction may span several orders of magnitude, as long as the dark matter is cold, massive and weakly interacting. It appears reasonable to impose a log-prior on both parameters. For definiteness we choose $m_{\rm DM}$ to lie in the range $1 \to 1000$~GeV and allow $\sigma_n^{\rm  SI}$ to vary between $10^{-46} \to 10^{-36}\ {\rm cm}^2$. For the parameters $\delta$ and $f_n/f_p$ there are no constraints without relying on a specific dark matter model. As a guidance we can use their definition: for instance for splitting $\delta$ bigger than 300 keV the scattering is highly suppressed because of kinematics, hence reasonably $\delta$ goes from 0 up to 300 keV. Exothermic dark matter favors light nuclei and light WIMPs, hence it is fair to consider $\delta$ negative starting from -100 keV. On the other hand the suppression of the scattering over neutron or proton as a function of $f_n/f_p$ becomes asymptotically flat for $f_n/f_p = -2$ and $f_n/f_p=1$, which are then the two extrema $\theta_{\rm min}$ and $\theta_{\rm max}$ for the prior pdf on the isospin violating parameter. The astrophysical parameters follow Gaussian priors centered on their most probable values.

Provided the data are sufficiently constraining the marginal posterior typically exhibits very little dependence on the choice of prior~\footnote{This occurs if the prior pdf is nearly constant and, under a parameter transformation  $f: \theta \to \theta'$, the mapping $f$ is almost linear over the parameter region where the likelihood is large.}. For data that can only provide an upper or a lower bound on a parameter (or no bound at all) however, the properties of the inferred posterior and the boundaries of credible regions can vary significantly with the choice of prior as well as its limits $\theta_{\rm min}$ and $\theta_{\rm max}$, making an objective interpretation of the
results rather difficult.  It has been proposed in~\cite{Arina:2011si} an alternative to computing credible intervals from the fractional volume of the marginal posterior in the  \{$m_{\rm DM},\sigma_n^{\rm SI}$\}-subspace $\mathcal{P}_{\rm marg}(m_{\rm DM},\sigma_n^{\rm SI}|d)$. This one consists in constructing intervals based on the volume of the marginal posterior in $S$-space $\mathcal{P}_{\rm marg}(S|d)$, where $S$ is the expected WIMP signal, using a uniform prior on $S$ with a lower boundary at zero~\cite{Helene:1982pb}. An $x$\% upper bound thus constructed has a well-defined Bayesian interpretation that the probability of $S\leq S_x$ is $x$\%.  The limit $S_x$ is then mapped onto the  \{$m_{\rm DM},\sigma_n^{\rm SI}$\}-plane by identifying those combinations of  $m_{\rm DM}$ and $\sigma_n^{\rm SI}$ with $\mathcal{P}_{\rm marg}(m_{\rm DM},\sigma_n^{\rm SI}|d) =\mathcal{P}_{\rm marg}(S_x|d)$. A $x$\% contour computed in this manner has the property of being independent of our choice of prior boundaries for $m_{\rm DM}$  and $\sigma_n^{\rm SI}$.  Its drawback, however, is that it has no well-defined  probabilistic interpretation in \{$m_{\rm DM},\sigma_n^{\rm SI}$\}-space. To distinguish these $S$-based credible intervals from the conventional ones based on the volume of $\mathcal{P}_{\rm marg}(m_{\rm DM},\sigma_n^{\rm SI}|d)$, we denote them with a subscript ``$S$'', such as $90_S$\%. The prior range in inference problem can be updated, namely one can choose as prior the posterior pdf of a previous independent observation.

\section{Updated status of dark matter theoretical parameter inference}\label{sec:inf}
In this section we review Bayesian statistical inference for a selection of recent experimental results (XENON100, as representative of upper bound and DAMA, CRESST, CoGeNT and CDMS-Si) to summarize the status of current dark matter direct searches. The main results can be found in~\cite{Arina:2011si,Arina:2012dr}, however here we update the inference by including CDMS-Si data. The procedure can be applied similarly to all other upper limits. Of particular interest for light WIMPs are the exclusion bounds of COUPP, PICASSO and SIMPLE, bubble chambers made of light elements, $\rm CF_3I$, ${\rm C_4 F_{10}}$ and ${\rm C_2 Cl F_5}$ respectively, see~\cite{Arina:2012dr}.
We first discuss the dependence of the credible regions on the priors and nuisance parameters assuming elastic SI interaction, to exemplify the features of Bayesian statistical analysis.

\begin{figure}[t!]
\begin{minipage}[t]{0.49\textwidth}
\centering
\includegraphics[width=1.\columnwidth,trim=40mm 81mm 47mm 86mm,clip]{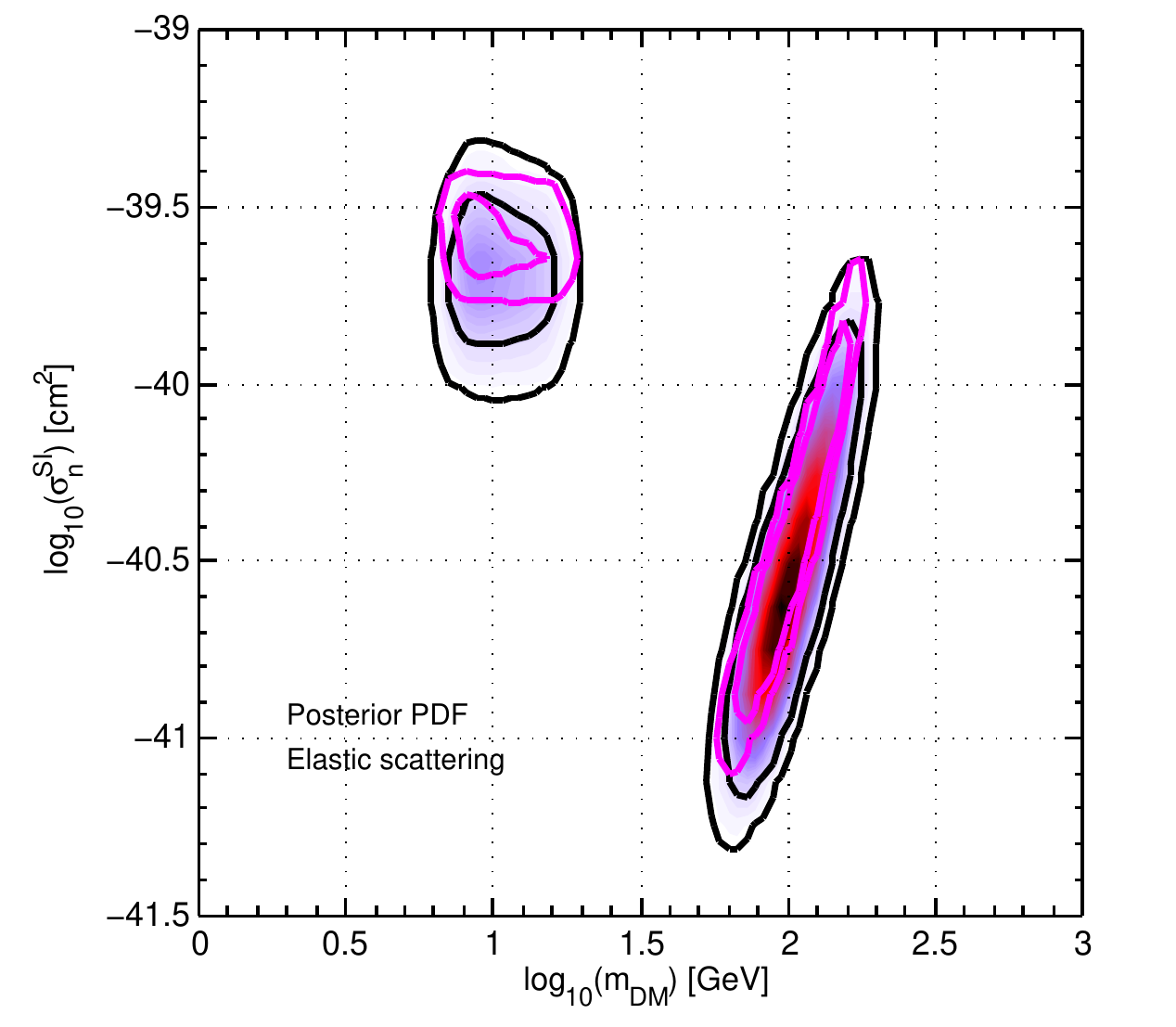}
\end{minipage}
\begin{minipage}[t]{0.49\textwidth}
\centering
\includegraphics[width=1.03\columnwidth,trim=47mm 86mm 47mm 92mm,clip]{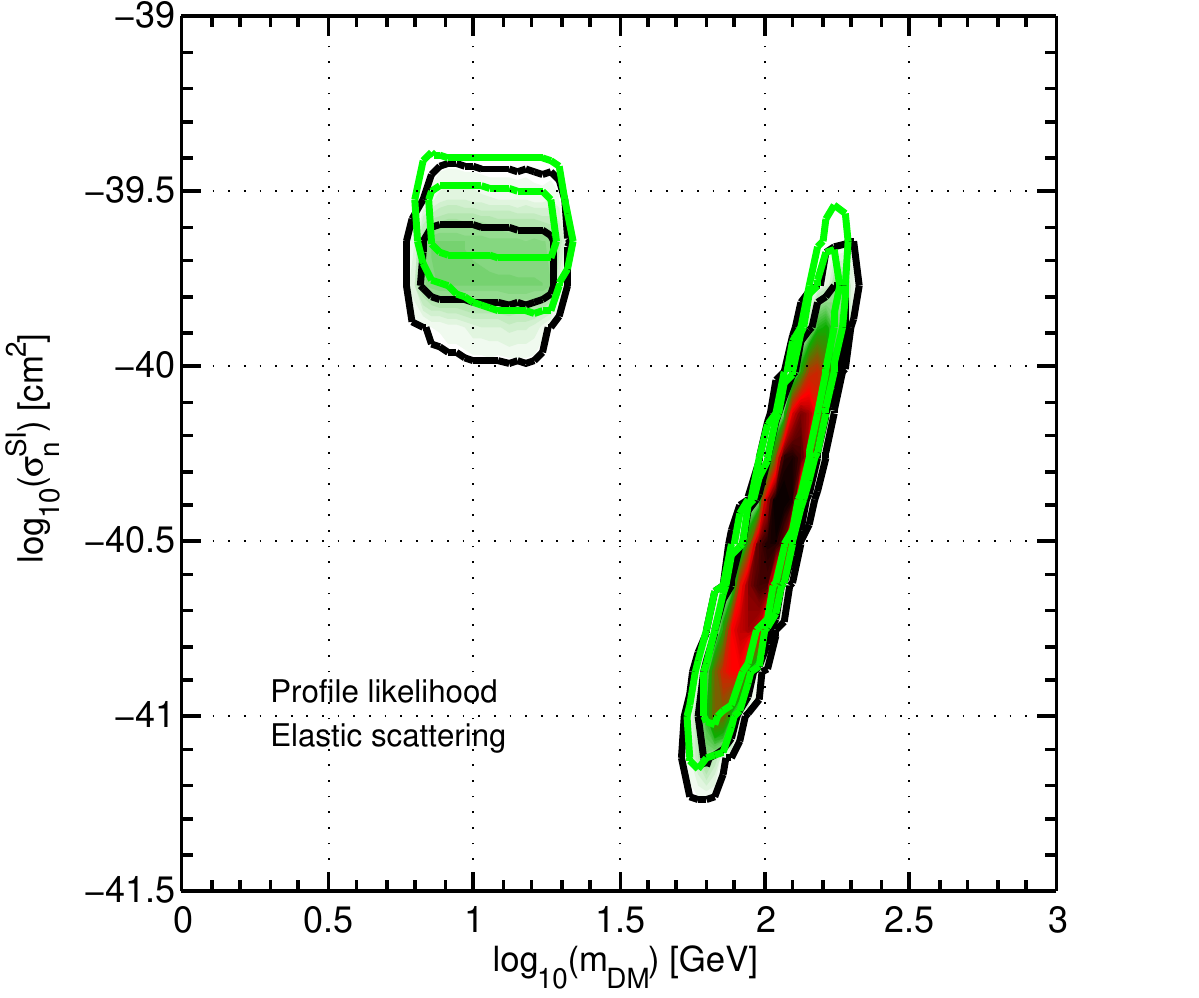}
\end{minipage}
\caption{Inference for DAMA. {\it Left:}  2D marginal posterior in the \{$m_{\rm DM},\sigma_n^{\rm SI}$\}-plane. The magenta (gray) lines enclose the credible regions assuming the SHM and fixed astrophysical variables, while the black lines stand for marginalized astrophysics (NFW dark matter density distribution). The lines enclose the 90\% and the 99\% credible regions in both cases. {\it Right:} Same as left for the profile likelihood: green (light gray) lines denote the SHM and the black contours with the shaded regions are for the marginalization over astrophysics. The solid contours correspond to $\Delta \chi^2_{\rm eff} = 4.6,9.2$. The color code in both cases goes from low to high posterior pdf (profile likelihood) values going from the light to dark color. Both posterior pdf and profile likelihood are normalized with respect to the maximum.}
\label{fig:dama}
\end{figure}

Let's start with the inference for DAMA, shown in Fig.~\ref{fig:dama}.  There are two quenching factors $q_{Na}$ and $q_{I}$ as nuisance parameters, because DAMA is a scintillator made by NaI crystals. The 2D marginal posterior pdf and the profile likelihood in the \{$m_{\rm DM}, \sigma_n^{\rm SI}$\}-subspace are shown by the solid magenta and green line in the left and right panel respectively, for the SHM but systematics integrated/profiled out. We remind that the detector can not disentangle if the recoil happened on Na or I, as a results there are two credible regions: the one at low mass is due to scattering off sodium, while the one at large $m_{\rm DM}$ is due to scattering off iodine. Both Bayesian and profile likelihood approaches single out the two preferred islands of parameter space in  \{$m_{\rm DM}, \sigma_n^{\rm SI}$\}, and indicate that $\mathcal{P}_{\rm marg}(m_{\rm DM},\sigma_n^{\rm SI})$ and $\mathcal{P}_{\rm prof}(m_{\rm DM},\sigma_n^{\rm SI})$
coincide to a good degree\footnote{For the profile likelihood, we show the two $\Delta \chi^2_{\rm eff}$ contours defined via likelihood ratio and with the value of $4.6, 9.2$ for the classical 90\% and 99\% confidence intervals for two degrees of freedom 
(assuming Wilks' theorem holds).}, meaning that the nuisance directions contribute no strong volume effects. This agreement indicates that when the data are sufficiently informative  so that the likelihood function overcomes the dependence on the priors, Bayesian and classical statistical methods yield very similar inference results. However both quenching factors show a flat 1D marginal posterior and profile likelihood (Fig.~\ref{fig:combin}, middle panel, light blue solid and dashed curve for $q_{\rm Na}$), meaning that ultimately the DAMA data do not constrain either $q_{Na}$ or $q_I$. 

The effect of including the astrophysical uncertainties is illustrated by the shaded region in both panels of Fig.~\ref{fig:dama}. From the posterior pdf we learn that there are volume effects due to the marginalization over these uncertainties. Conversely the profile likelihood is not significantly altered by the volume of the astrophysical uncertainties, as expected by definition of likelihood ratio. In general all dark matter density profiles give very similar inference results (here we assumed NFW density profile, see~\cite{Arina:2011si} for details). This means that the exact shape of the dark matter halo density profile, within the class of spherically symmetric, smooth profiles, does not yet play a role in direct searches. This conclusion is further supported by the inferred local dark matter density, circular and escape velocities. The preferred values for these quantities differ from profile to profile however, once the dark matter halo profile has been fixed, the preferred values for $v_0$, $v_{\rm esc}$ and $\rho_{\odot}$ and their associated uncertainties are virtually independent of the additional constraints from the experiments. WIMP direct searches are not at the moment contributing towards constraining the astrophysics of the problem and this is the reason why $\mathcal{P}_{\rm marg}(m_{\rm DM},\sigma_n^{\rm SI})$ exhibits volume effects. The same reasoning as above applies to inference of CoGeNT and CRESST data.

\begin{figure}[t!]
\begin{minipage}[t]{0.32\textwidth}
\centering
\includegraphics[width=1.\columnwidth,trim=22mm 27mm 13mm 13mm,clip]{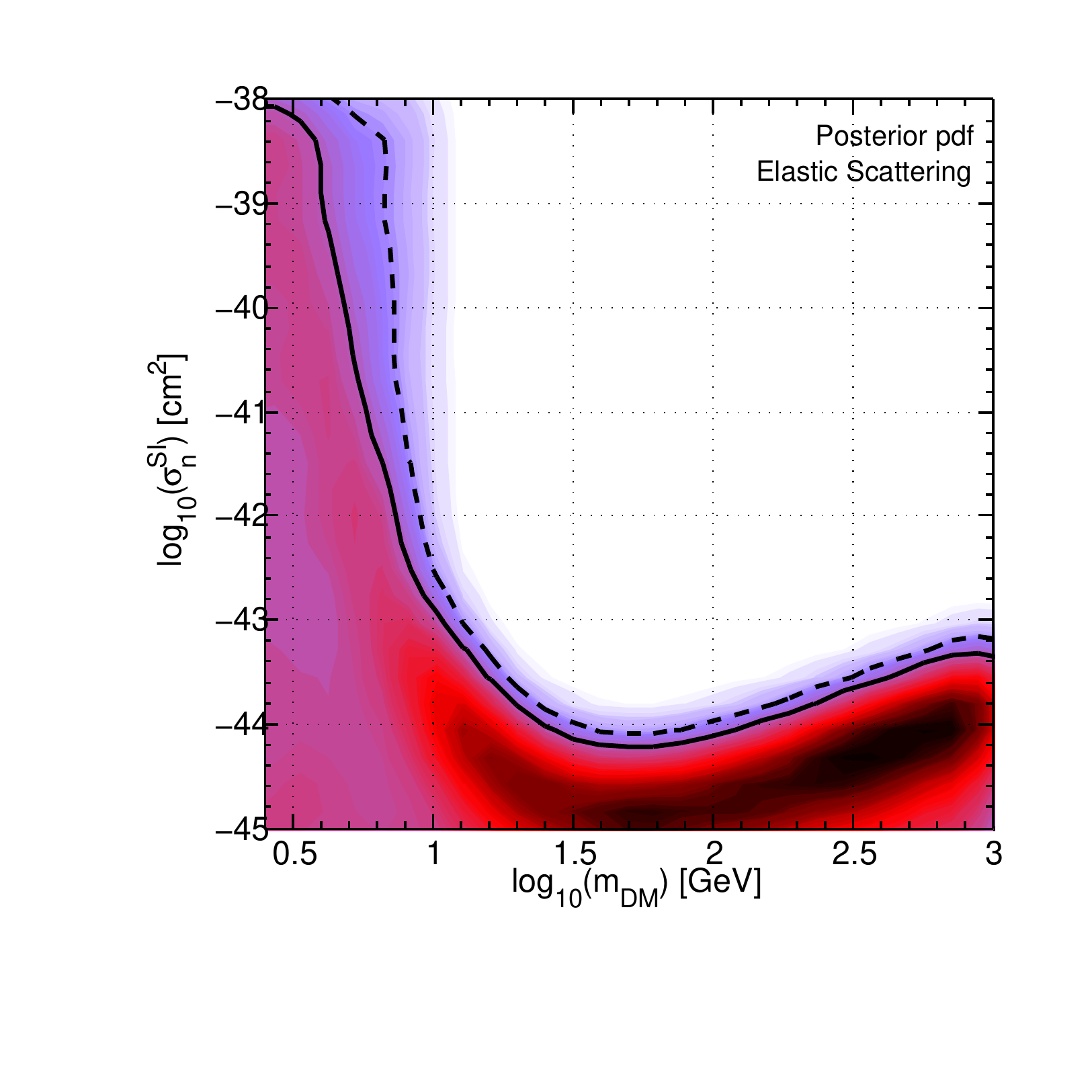}
\end{minipage}
\begin{minipage}[t]{0.32\textwidth}
\centering
\includegraphics[width=1.\columnwidth,trim=44mm 92mm 53mm 95mm,clip]{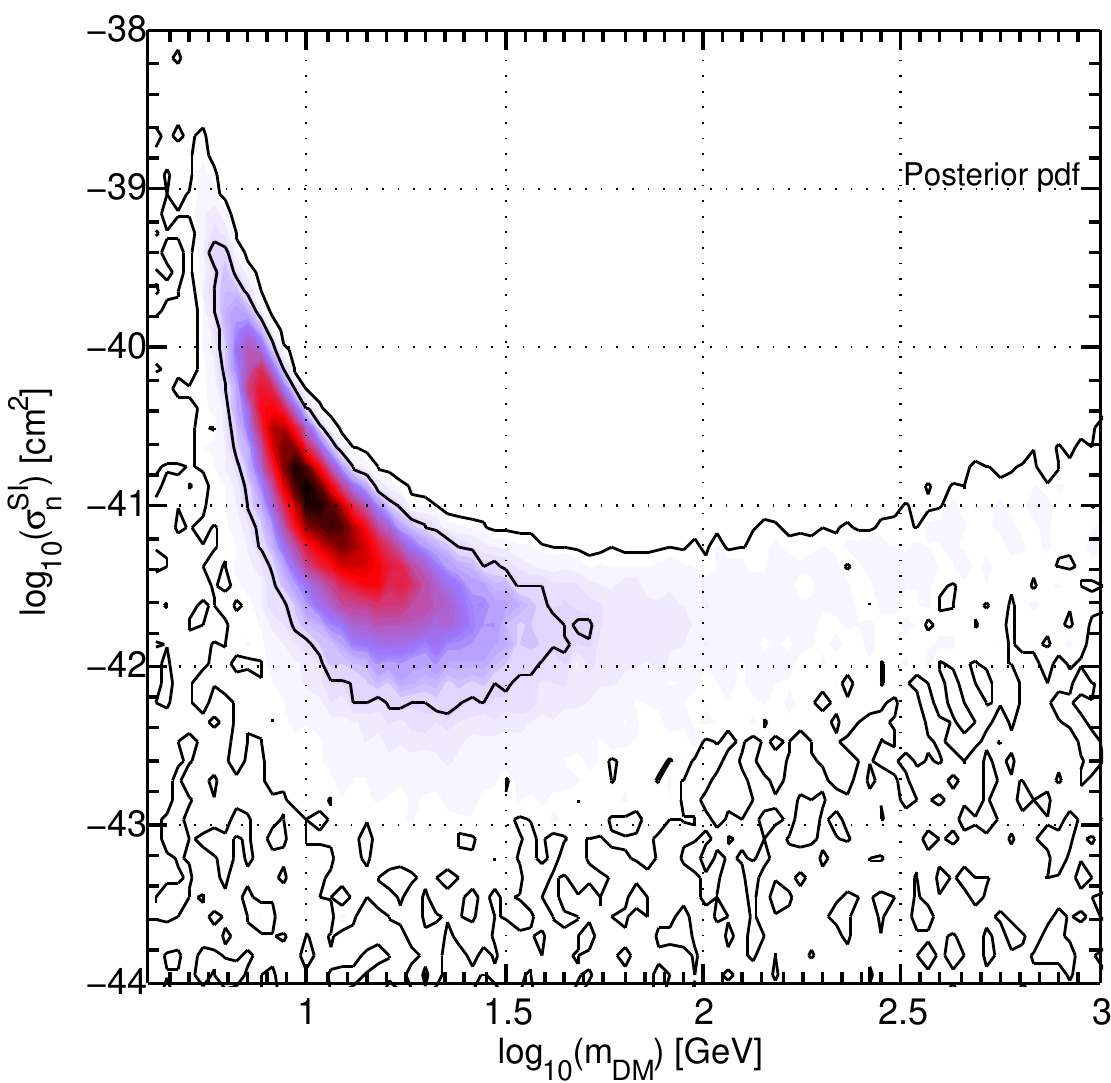}
\end{minipage}
\begin{minipage}[t]{0.32\textwidth}
\centering
\includegraphics[width=1.\columnwidth,trim=44mm 92mm 53mm 95mm,clip]{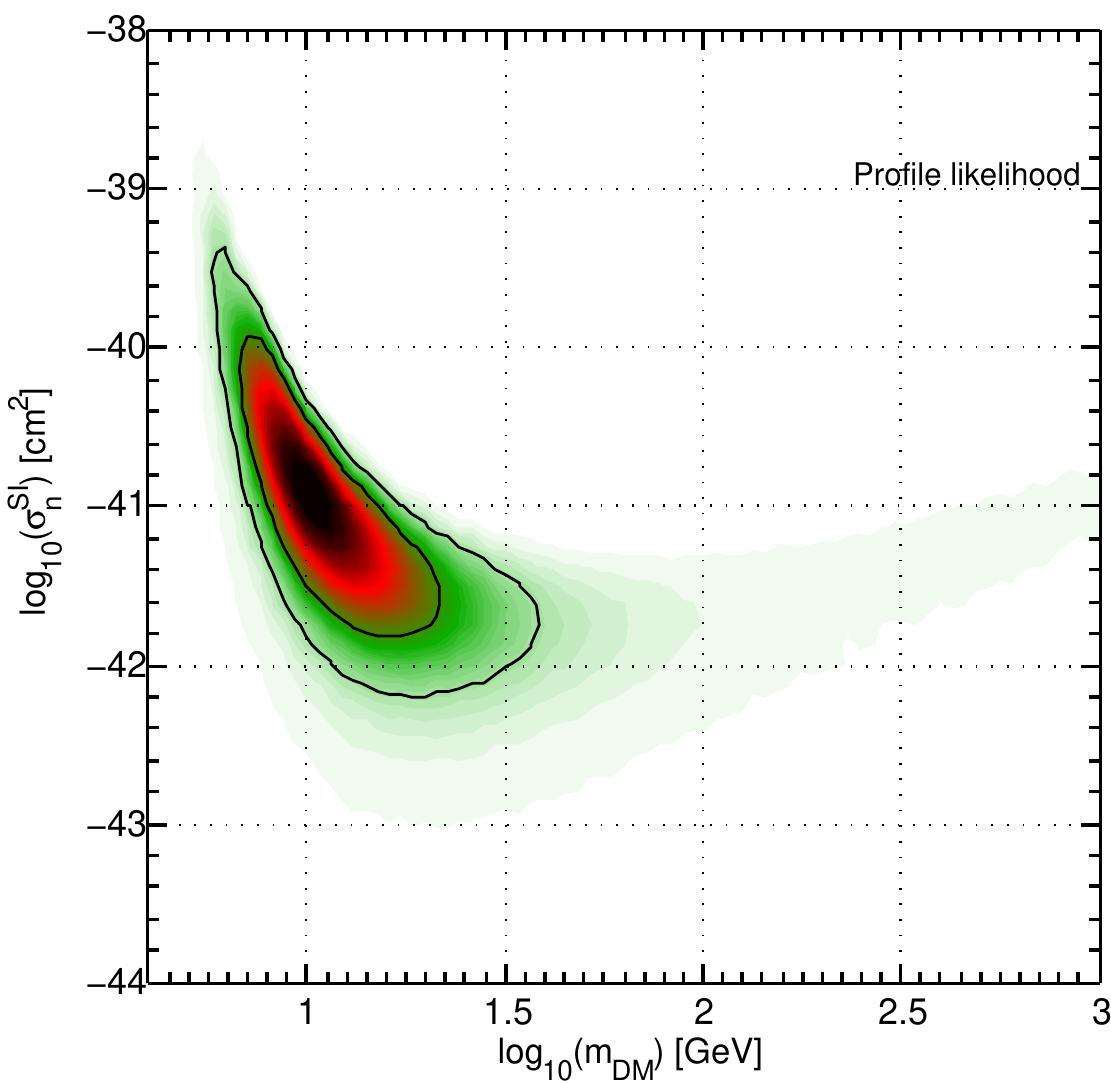}
\end{minipage}
\caption{Inference for XENON100 and CDMS-Si. {\it Left:}  2D marginal posterior in the \{$m_{\rm DM},\sigma_n^{\rm SI}$\}-plane for XENON100. The black line encloses the 90\% credible region, while the dashed line the $90_S\%$ corresponding to a $\Delta\chi^2_{\rm eff} = 3.1$ (see~\cite{Arina:2011si,Arina:2012dr}). {\it Center:} 2D marginal posterior in the \{$m_{\rm DM},\sigma_n^{\rm SI}$\}-plane for CDMS-Si. The black lines  enclose the 68\% and the 90\% credible regions. {\it Right:} Same as the central panel for the profile likelihood and the solid contours correspond to $\Delta \chi^2_{\rm eff} = 2.3,4.6$. The systematics are marginalized over and the SHM is assumed. Color coding as in figure~\ref{fig:dama}.}
\label{fig:explain}
\end{figure}

The case of CDMS-Si is different, as illustrated in Fig.~\ref{fig:explain}, center and right panels. Here the contours denote the 68\% and 90\% credible regions and  $\Delta \chi^2_{\rm eff}=2.3,4.6$ in the \{$m_{\rm DM}, \sigma_n^{\rm SI}$\}-plane, for $\mathcal{P}_{\rm marg}(m_{\rm DM},\sigma_n^{\rm SI})$ and $\mathcal{P}_{\rm prof}(m_{\rm DM},\sigma_n^{\rm SI})$ respectively\footnote{The small blobs in $\mathcal{P}_{\rm marg}(m_{\rm DM},\sigma_n^{\rm SI})$ do not have a physical meaning but are artifact due to the sampling of a very flat likelihood.}. These two quantities do not coincide meaning that  the data are not constraining enough to overcome the dependence on the prior in the marginal posterior pdf. The Bayesian intervals account for the uncertainties in the measurements and systematics and denote a more robust approach: the $90\%$ contour does not close, meaning that the tension claimed with the exclusion bound is not present. On the other hand the profile likelihood gives a measure of goodness of fit and is more constraining: the best fit point is $m_{\rm DM} = 8.6$ GeV and $\sigma^{SI}_n=10^{-40}\rm cm^2$, while the nuisance parameter $N_e$, which is the normalization of electron background as described in~\ref{sec:app}, peaks at 0.4: these values are compatible with the other analyses~\cite{Agnese:2013rvf,Frandsen:2013cna,DelNobile:2013cta}. In the following, we stick with Bayesian intervals and only show the 68\% to avoid cluttering.

Lastly we comment on the exclusion bound of XENON100, whose inference is shown Fig.~\ref{fig:explain}, left panel. Note that the 2D marginal posterior 
forms a plateau as $m_{\rm DM}$ and $\sigma_n^{\rm SI}$ approach their respective boundaries.  In this case, credible regions constructed from the
volume of the marginal posterior in  \{$m_{\rm DM}, \sigma_n^{\rm SI}$\}-space can be strongly dependent on the choice of the $m_{\rm DM}$ and $\sigma_n^{\rm SI}$ prior boundaries. On the other hand, the $90_S$\% bound (black dashed line) is  independent of the boundary conditions as discussed in Sec.~\ref{sec:dp} (see~\cite{Arina:2011si} for details) and is used in the following discussions\footnote{Following the color code the excluded region lies in the right-hand side of the $90_S\%$ credible contour.}. This exclusion limit on $\sigma_n^{\rm SI}$ at low WIMP masses ($m_{\rm DM} \leq 30$~GeV) agrees with the one provided by XENON100~\cite{Aprile:2012nq}, while at high masses it is slightly less constraining because of the approximated likelihood we are using with respect to the one of the collaboration. The marginalization over astrophysical uncertainties has the effect of shifting towards the right-hand side the exclusion bounds in the \{$m_{\rm DM}, \sigma_n^{\rm SI}$\}-subspace, slightly weakening the tension between experimental results. We do not comment any further on astrophysical uncertainties and present the results marginalizing over them. We conclude this overview on the effect of prior pdfs on the direct detection data noticing that the choice of prior boundaries on $m_{\rm DM}$ and $\sigma_n^{\rm SI}$ translates directly to how likely we deem the direct detection experiments to actually make a positive detection. Consider the loss of detection sensitivity for large masses (due to the large mass splitting between the dark matter particle and the nucleus), or for light WIMPs (because of the energy threshold):  the larger the prior-space in the \{$m_{\rm  DM},\sigma_n^{\rm SI}$\}-plane, the smaller the relative fraction that the experiments will be able to probe, and the smaller the subjective prior probability for them to see something. 

\begin{figure}[t!]
\begin{minipage}[t]{0.49\textwidth}
\centering
\includegraphics[width=1.\columnwidth,trim=43mm 83mm 47mm 87mm,clip]{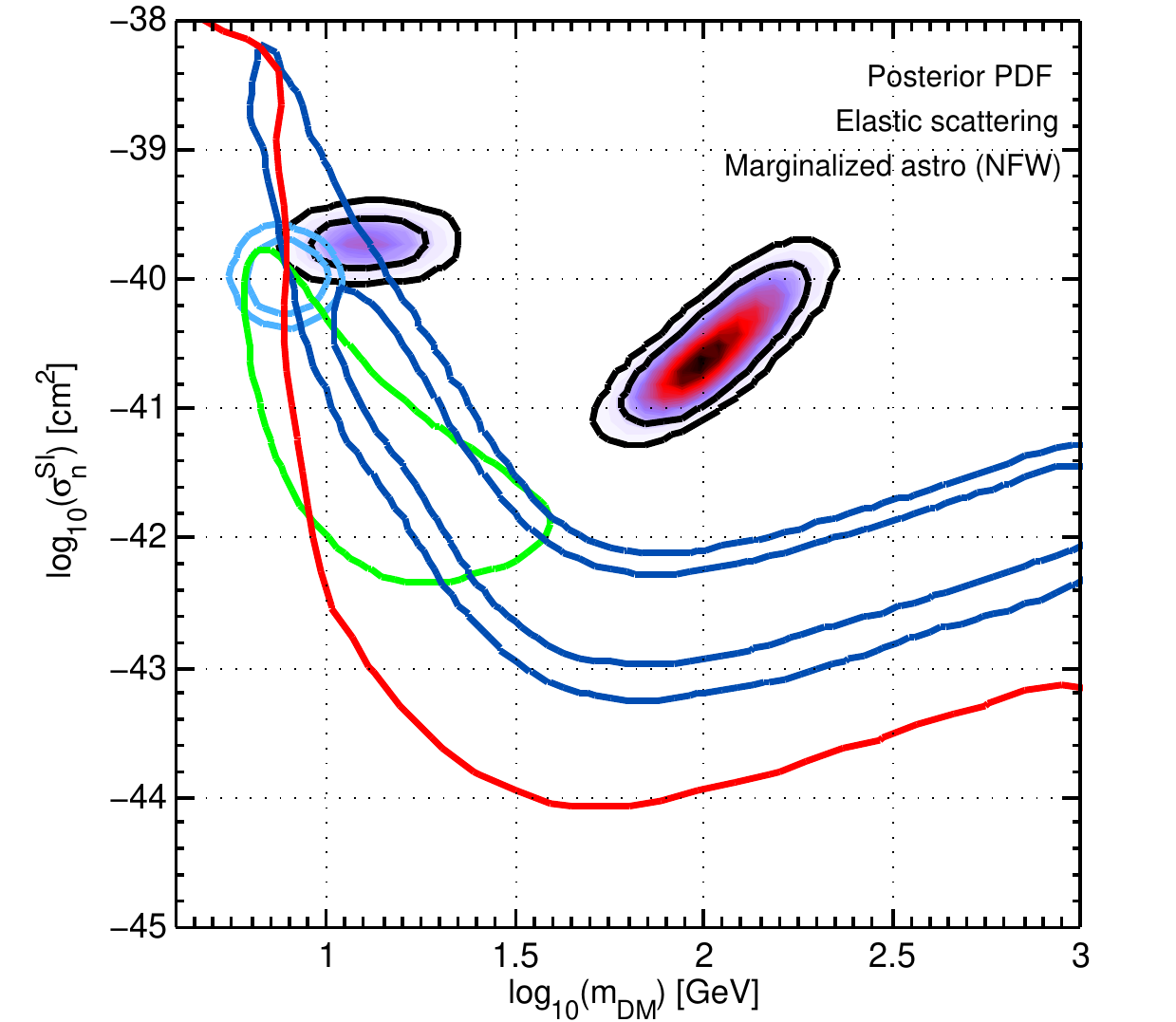}
\end{minipage}
\begin{minipage}[t]{0.49\textwidth}
\centering
\includegraphics[width=1.\columnwidth,trim=43mm 83mm 47mm 87mm,clip]{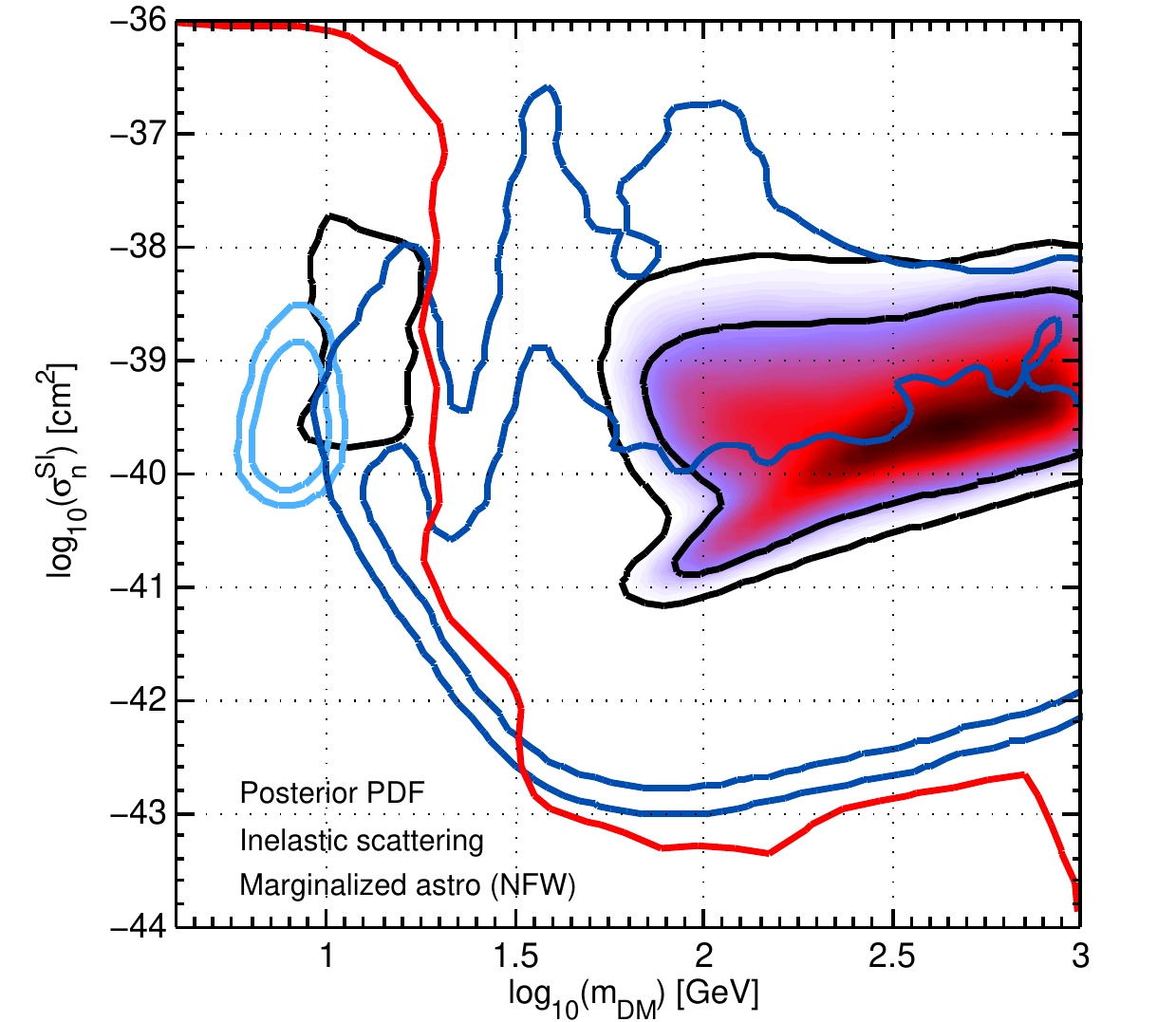}
\end{minipage}
\\
\hspace*{-0.2cm}			
\begin{minipage}[t]{0.49\textwidth}
\centering
\includegraphics[width=1.\columnwidth,trim=43mm 83mm 47mm 87mm,clip]{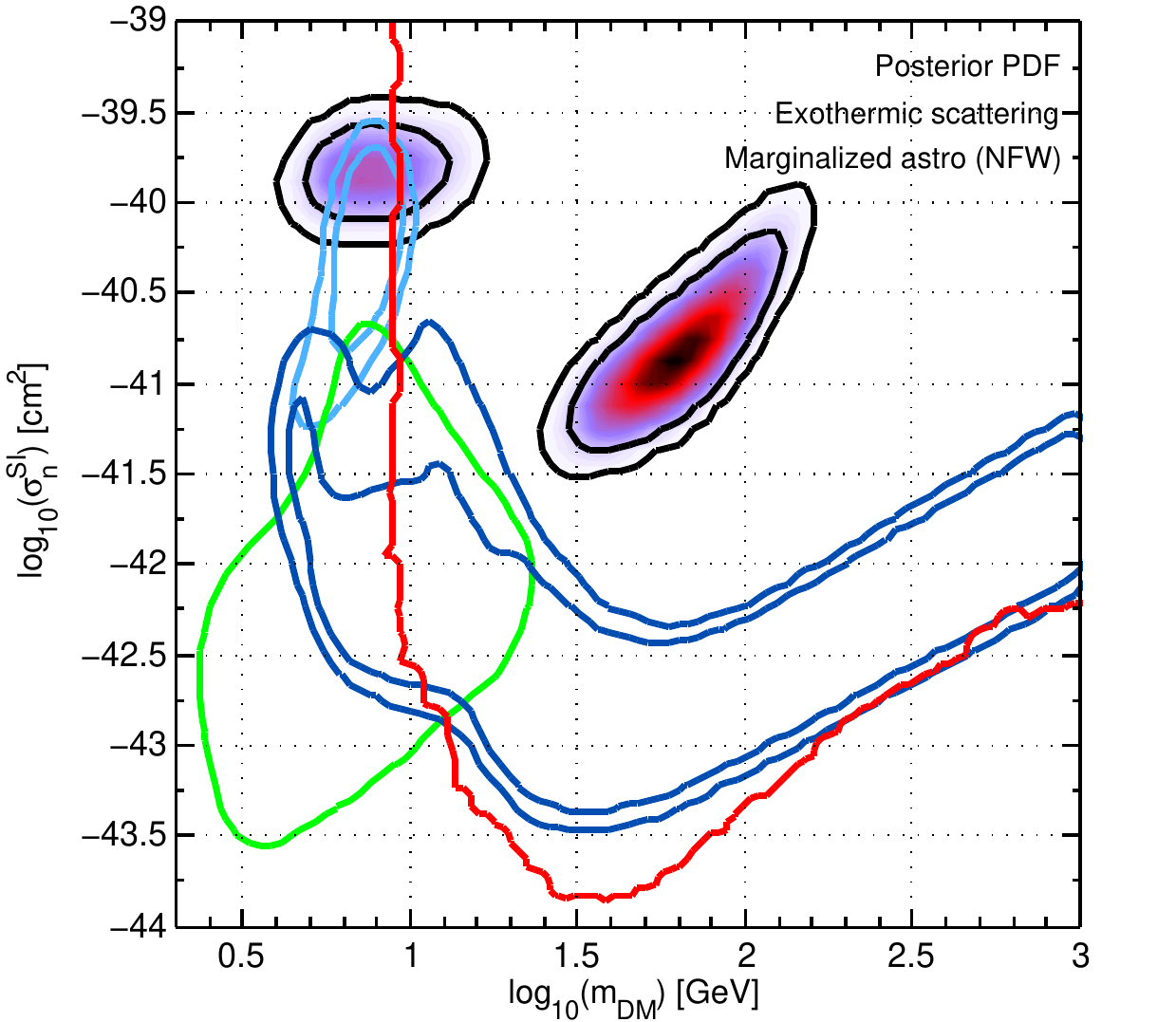}
\end{minipage}
\begin{minipage}[t]{0.49\textwidth}
\centering
\includegraphics[width=1.\columnwidth,trim=43mm 83mm 47mm 87mm,clip]{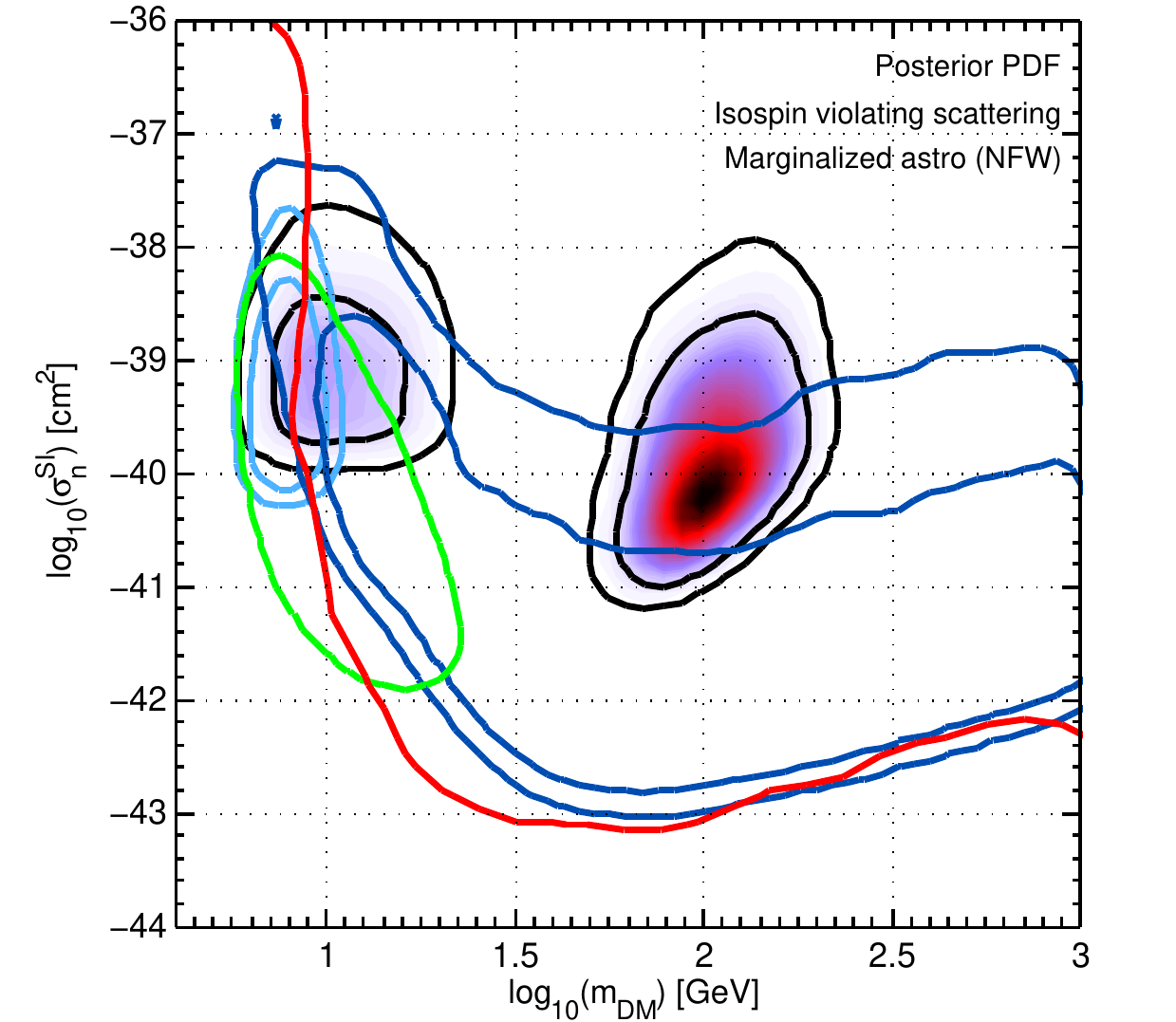}
\end{minipage}
\caption{2D credible regions for the individual experimental bounds and marginalized astrophysics (NFW dark matter density distribution), combined in a single plot. For DAMA (shaded), CoGeNT (light blue) and CRESST (blue) we show the 90\% and 99\% contours, while for CDMS-Si (green) only the 68\% to avoid cluttering. The red solid line represents the $90_S$\% bound for XENON100 corresponding to $\Delta\chi^2_{\rm eff} = 3.1$. {\it Top Left and Right, Bottom Left and Right:} Elastic, inelastic,  exothermic and isospin violating scattering scenario as labelled. Color coding as in figure~\ref{fig:dama}.}
\label{fig:soa}
\end{figure}

We review the Bayesian inference for the particle physics models considered here in Fig.~\ref{fig:soa}, which displays the 2D marginal posterior pdf for each individual experiment in the \{$m_{\rm DM}, \sigma_n^{\rm SI}$\}-space, plotted all together. We comment on the salient features and discrepancies with respect to classical statistical analysis.
\paragraph {Elastic SI scattering (top left)} 
The CoGeNT region is denoted by the light blue contours at 90\% and 99\%: even thought the detector is made by Ge, the preferred WIMP mass is $\sim 7$ GeV because of the very low experimental threshold. CRESST is a scintillator made by $\rm CaWO_4$ and it is sensitive to low (because of O), medium (because of Ca) and high mass (due to W). The parameter regions favored by DAMA and CRESST are only very marginally compatible with the $90_S\%$ credible regions of XENON100. In contrast, using the Bayesian credible intervals marginalized over all uncertainties, the compatibility between CDMS-Si and XENON100 increases with respect to the classical statistical analysis~\cite{Agnese:2013rvf,Frandsen:2013cna,DelNobile:2013cta}. While the XENON100 collaboration claims that their exclusion limit has ruled out the CoGeNT preferred region~\cite{Aprile:2012nq}, we have found that when systematics, such as the scintillation efficiency ${\rm L}_{\rm eff}$ at low recoil energies, and the astrophysical uncertainties are accounted for, the CoGeNT and the XENON100 data can find some common ground. Between CoGeNT, DAMA, CRESST and CDMS-Si we have found that their 99\% credible regions at least overlap. This is a consequence of the choice of prior boundaries: for instance we used for $q_{\rm Na}$  the range $0.2 \to 0.6$ exploiting the correlation $m_{\rm DM}-q_{\rm Na}$ for which the larger the quenching factor the smaller the WIMP mass\footnote{This choice of $\theta_{\rm min}$ and $\theta_{\rm max}$ encompasses the whole experimental range, however the most recent measurements favor the lower values, see~\cite{Arina:2011si} for details.}.  It is however  important to examine the degree of overlap between the preferred regions in the other parameter directions, as we discuss later.
\paragraph{Inelastic SI scattering (top right)} 
This plot (and the following as well) shows the impact of marginalizing over the extra free theoretical parameter, a novelty introduced in~\cite{Arina:2012dr}\footnote{Usually several 2D plots are provided, each at a fixed value of the parameter. Same holds for the astrophysical parameters or other nuisances.}. Note the huge volume effects because all direct detection data are not constraining enough to support the new free parameters $\delta$. The DAMA favored region is the one on iodine, because the interaction favors heavy nuclei, while the scattering off Na nuclei is present only in the 99\% volume space of the posterior pdf.  The well known tension between XENON100 and the DAMA iodine region is retrieved. CRESST and CDMS-Si (which is not shown to avoid cluttering) suffer the most of volume effects, leading to wide credible regions. CoGeNT and DAMA on Na are compatible with XENON100 bound, which is weakened by the choice of the inelastic interaction. However note that the detection regions at very light WIMP mass are challenged by the exclusion limits of PICASSO or SIMPLE~\cite{Arina:2012dr}.
\paragraph{Exothermic SI scattering (bottom left)}
Note that this extra parameter receives more support from the data than in the previous inelastic case, as the volume effects are less prominent. We have found that this particle physics scenario accommodates CDMS-Si and XENON100 at $90_S\%$, result compatible with~\cite{Frandsen:2013cna,DelNobile:2013cta}. Moreover we have found that all the other detection regions are compatible at $90_S\%$ with XENON100. We discuss in the model comparison section the poor compatibility between detection regions.
\paragraph{Isospin violating scattering (bottom right)}
This scenario seems to be the one that releases the most the tension between all experimental results, even though the volume effects due to $f_n/f_p$ are significant. The additional parameter $f_n/f_p$ is responsible for the increased compatibility below a WIMP mass of 9 GeV: the scattering off Xe is suppressed with respect to the scattering off Na or Si (and partially off Ge) for value of $f_n/f_p\sim -0.7$. This behavior is confirmed by the 1D marginal posterior pdf for the extra parameter in Fig.~\ref{fig:1dpar} (see also~\cite{Arina:2012dr} for details).  The outcome of this Bayesian analysis has been shown to be compatible with the particular cases presented in~\cite{Feng:2011vu,Feng:2013vod}. 

The same Bayesian statistical procedure is applicable to spin-dependent scattering as well, provided that it is possible to write a likelihood function for the experiment. For instance this is the case of PICASSO or SIMPLE experiments, see~\cite{Arina:2012dr} for details. Regarding a standard elastic spin-dependent interaction the inclusion of experimental and  astrophysical uncertainties will have an impact analogous to the case of SI independent interaction discussed here. Indeed the exclusion bounds are less tight and the detection regions are wider  because of volume effects. The tension between experimental results will still be present. Considering a frequentist approach, one of the most recent update on spin-dependent interaction can be found in~\cite{Buckley:2013gjo}. A Bayesian analysis of non-standard interactions has not be done yet, and in this case, as there is a additional velocity and/or momentum dependence on the differential rate, a rigorous treatment of astrophysical uncertainties and dark matter halo models might reveal interesting.

\section{The power of Bayesian evidence in model comparison}\label{sec:idc}
In dark matter direct searches, Bayesian model comparison has been used in~\cite{Arina:2011zh} to address the question whether the annual modulated signal claimed by CoGeNT is due to dark matter  or to other effect, and in~\cite{Arina:2012dr} to search for the best particle physics scenario that accommodates the experimental results. In the following we review these analyses and summarize the main findings.

\subsection{Do CoGeNT data provide a hint for WIMP annual modulation?}\label{sec:cogm}
The CoGeNT collaboration has reported a modulating signal in the energy range $0.5 \to 3.0$~keVee at $2.8\sigma$, with a modulation amplitude of $16.6 \pm 3.8 \%$, a period of $347 \pm 29$~days, and the minimum rate falling on October~16$\pm 12$~days~\cite{Aalseth:2011wp}, somewhat at odds with the dark matter prediction of December 2. We review how to evaluate in a quantitative manner the probability of various explanations ({\it i.e.} models) for the presence (or absence) of modulation in the CoGeNT time-dependent data, and in particular to estimate the probability of the data being due to a light mass WIMP signal (the detailed analysis is done in~\cite{Arina:2011zh}). To this end, we adopt a phenomenological approach, and by virtue of Eq.~(\ref{eq:mod}), parameterize the time-dependent event rate $R(t)$ in the energy bin as 
\begin{equation}
\label{eq:fit}
R_i(t) = U^i_{\rm m} \left(1 + S^i_{\rm m} \cos[2 \pi (t-t_{\rm max} - 28)/T]\right)\,,
\end{equation}
where $t$ is in units of days,  with $t=0$ corresponding to the first day of data-taking, {\it i.e.} December 4, 2009, $t_{\rm max}$ is the phase of the modulation in terms of days since January 1, $T$ the modulation period and $i=1,2,3$ denotes the number of energy bins. Two more parameters, $U^i_{\rm m}$ and $S^i_{\rm m}$, denote the mean event rate and the fractional modulation respectively, with the superscript $i$ indicating that these quantities are generally energy-dependent. We treat all or subsets of $\{U^i_{\rm m}, S^i_{\rm m},t_{\rm max},T\}$ as free parameters, and determine using Bayesian model comparison if the CoGeNT data corroborate the dark matter hypothesis or some other alternative scenarios as follows.
\begin{enumerate}
\item No modulation (model 0): all parameters in Eq.~(\ref{eq:fit}) are set to 0 but the amplitudes $U^i_{\rm m}$. This is the reference and also simplest model
\item Modulation due to dark matter (model $1a$ for phenomenological assumption, model $1b$ for consistent signal). By assuming the modulation due to dark matter, the phase and period are fixed by the theoretical predictions hence the free parameters are the percentage of modulation $S^i_{\rm m}$ in addition to $U^i_{\rm m}$. What distinguish between $1a$ and $1b$ is the choice of prior pdf, in the first case it follows a uniform prior, while in the second case it is derived consistently from the total rate. WIMP theoretical predictions state that the modulation should be absent in the third energy bin.
\item Modulation due to some other physics (model $2a$: non dark matter but annual, model $2b$: non dark matter and free period). In addition to the free parameters already varied in the models 1, the phase in $2a$ follows a uniform prior, while in $2b$ the period as well can vary.
\end{enumerate}
\begin{figure}[t!]
\begin{minipage}[t]{0.49\textwidth}
\centering
\includegraphics[width=1.\columnwidth,trim=0mm 0mm 0mm 0mm,clip]{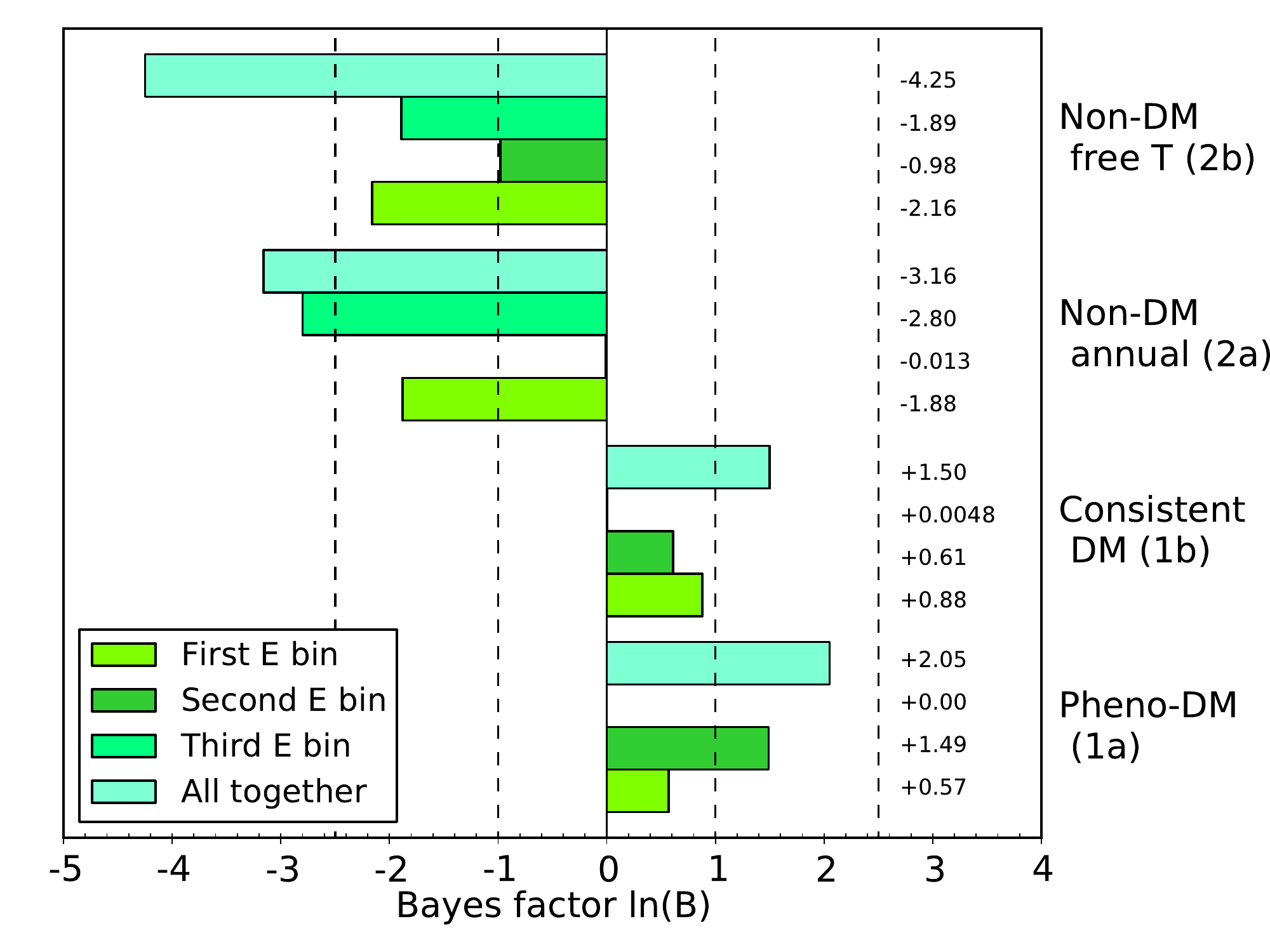}
\end{minipage}
\begin{minipage}[t]{0.49\textwidth}
\centering
\includegraphics[width=1.\columnwidth,trim=0mm 0mm 0mm 0mm,clip]{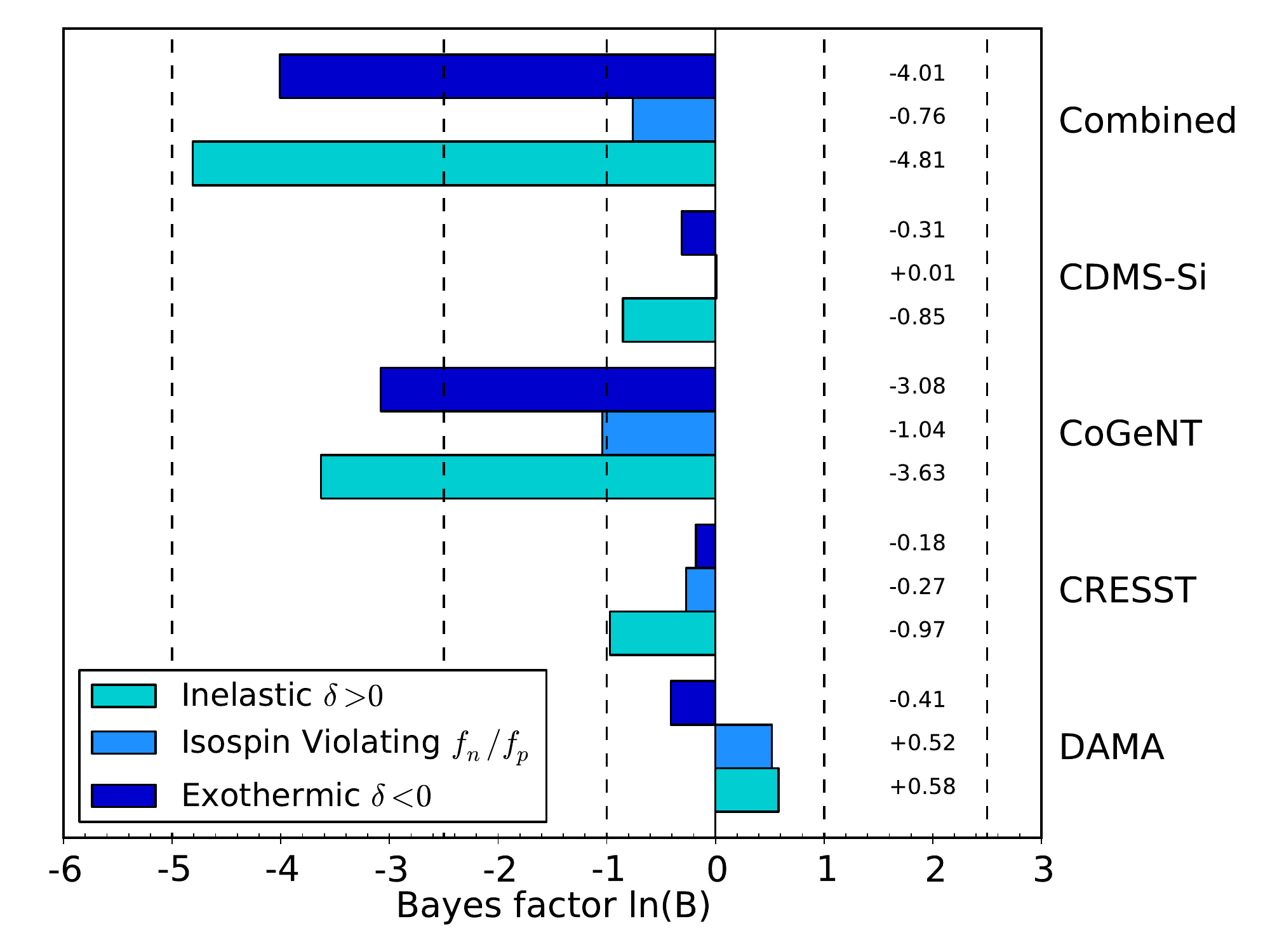}
\end{minipage}
\caption{Plot for the Bayes factor for the two model comparison cases described in the text. {\it Left:} Modulation models presented in~\ref{sec:cogm}. The models are specified on the vertical axis, while the different colors refer to the energy bin(s) for which the Bayes factors have been computed, as labelled in the plot. {\it Right:} Particle physics scenarios considered in Sec.~\ref{sec:gf}, analysis for the single experiments and the combined fit. The experiments are specified on the vertical axis, while the different colors refer to the model for which the Bayes factors have been computed, as labelled in the plot. The actual value of $\ln B$ in each case is indicated by the number in the right column and has an uncertainty of $\sim 0.02$ for the individual bins/experiments and $\sim 0.04$ for the combined analysis. Following Jeffrey's scale in Tab.~\ref{tab:jef}, the vertical lines demarcate the different empirical gradings of the strength of the evidence and the strength of evidence is computed with respect to model 0. The astrophysical uncertainties are marginalized over.}
\label{fig:lnbf}
\end{figure}
The results for the Bayes factor are shown in Fig.~\ref{fig:lnbf}, left panel, where the different colors correspond to the different energy bins and the combined analysis, as labelled. In the first energy bin  the evidence for modulation is inconclusive for dark matter models (models 1), and weakly against other physics models (models~2), when compared with the no modulation model. This yields a moderate to strong evidence for dark matter models as compared to the other physics models. In the second energy bin, compared with model 0, the support for modulation is at best weak.  With the exception of the moderate evidence for model $1a$ against model $2b$, the comparisons between the modulation models themselves are likewise weak to inconclusive. As far as it concerns the third bin, the CoGeNT data do not support the presence of modulation. The outcome of the combined analysis shows that the dark matter models $1a$ and $1b$ receive weak support from the data over no modulation. In contrast, models $2a$ and $2b$ are moderately disfavored with respect to the no modulation scenario. Dark matter models are thus strongly favored over other physics models, with odds in favor of the dark matter models ranging from $185:1$ to $560:1$. This is a consequence of the predictiveness of models $1a$ and $1b$: Occams' razor is at work, penalizing the other physics models for their excessive  free parameters unsupported by the data. Evidence (odds) between any pair of models can be obtained simply by adding (multiplying) the $\ln B$ (odds) reported in Fig.~\ref{fig:lnbf}.

For inference, only the posterior pdf of the combined analysis is meaningful (in the other cases it is multimodal, something that it is not always accounted for in classical statistical analysis), from which it appears that CoGeNT data support a higher modulation ($S^i_{\rm m} \sim 20\%$) than what is consistent with the total rate ($S^i_{\rm m} \leq 10\% $) and that the preferred phase and period for the modulation are $t_{\rm max}= 104\pm 10$ days and $T= 344 \pm 22$ days respectively. This approach does not attempt to reconcile the discording results of DAMA and CoGeNT: indeed DAMA measures a phase which is compatible with the standard prediction while the phase measured by CoGeNT peaks around April and is off by more than a month with respect to the dark matter predictions. Efforts to reconcile these experimental results go for instance into the direction of favoring an anisotropic dark matter velocity distribution, see {\it e.g.}~\cite{Kopp:2011yr,Arina:2011zh}, which also adjusts the modulated amplitude in CoGeNT and accommodates total and modulated rate.

\subsection{Combined fit of all positive detection experiments}\label{sec:gf}
\begin{figure}[t!]
\begin{minipage}[t]{0.32\textwidth}
\centering
\includegraphics[width=1.04\columnwidth,trim=43mm 81mm 40mm 86mm,clip]{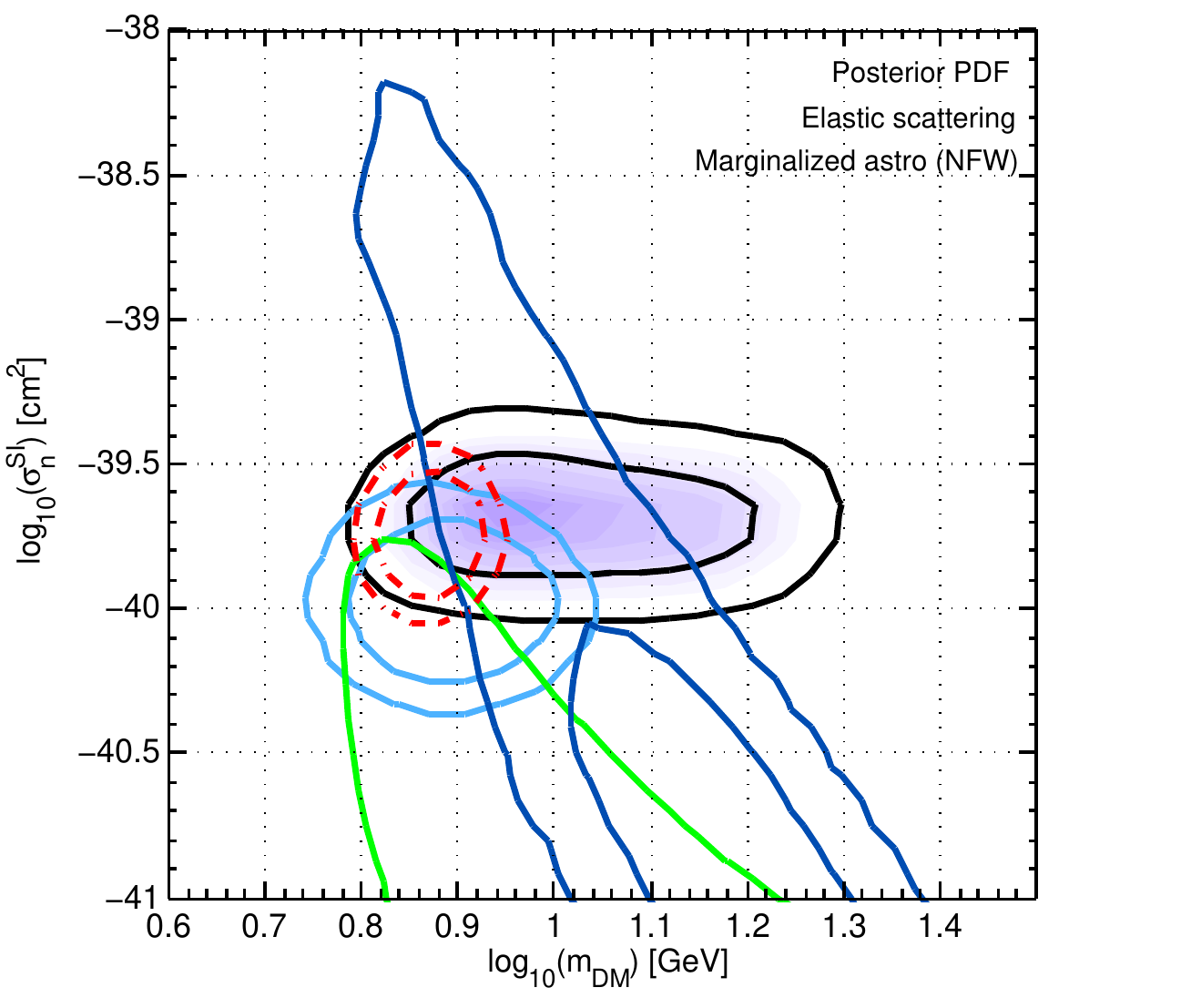}
\end{minipage}
\begin{minipage}[t]{0.32\textwidth}
\centering
\includegraphics[width=0.9\columnwidth,trim=25mm 25mm 10mm 13mm,clip]{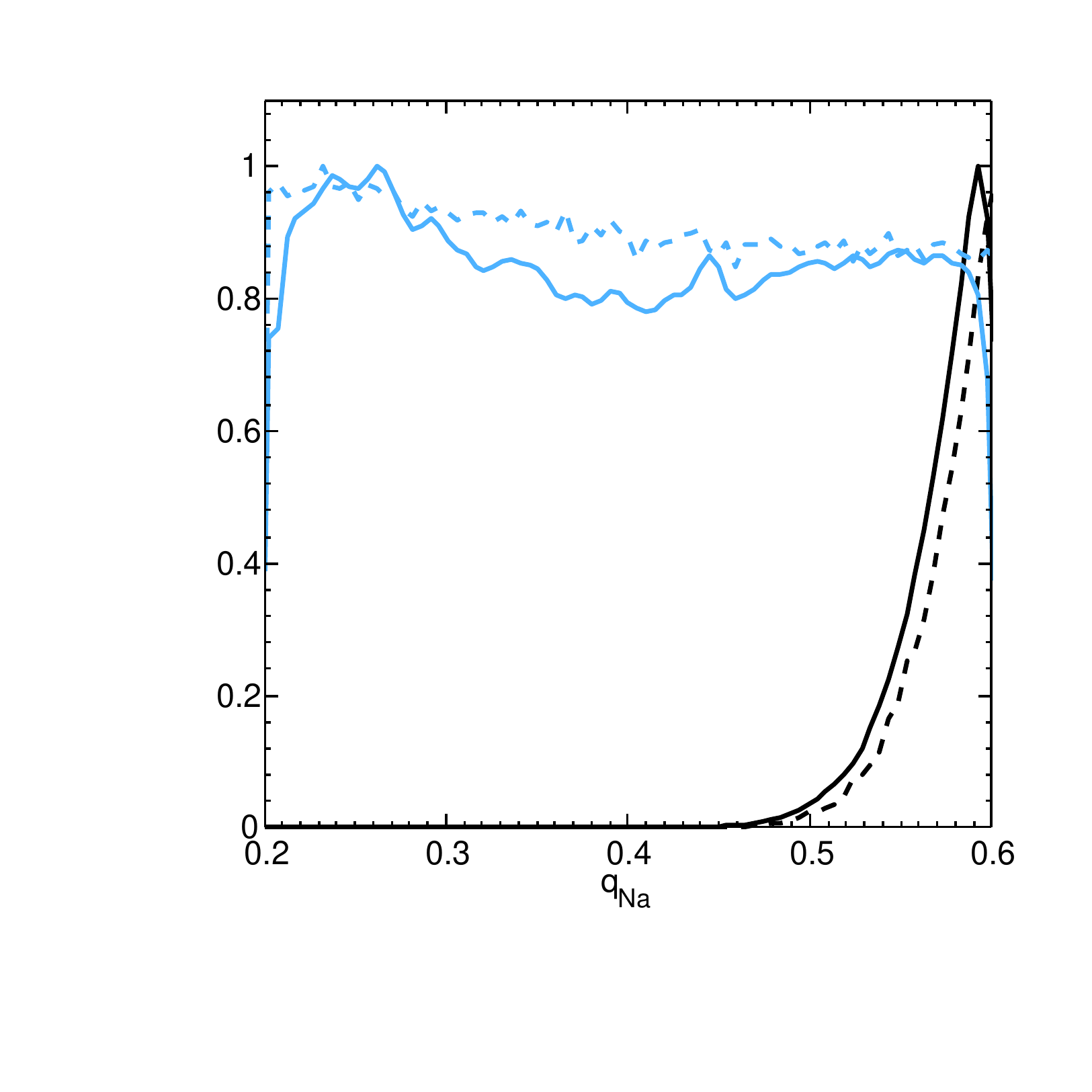}
\end{minipage}
\begin{minipage}[t]{0.32\textwidth}
\centering
\includegraphics[width=1.19\columnwidth,trim=32mm 94mm 40mm 100mm,clip]{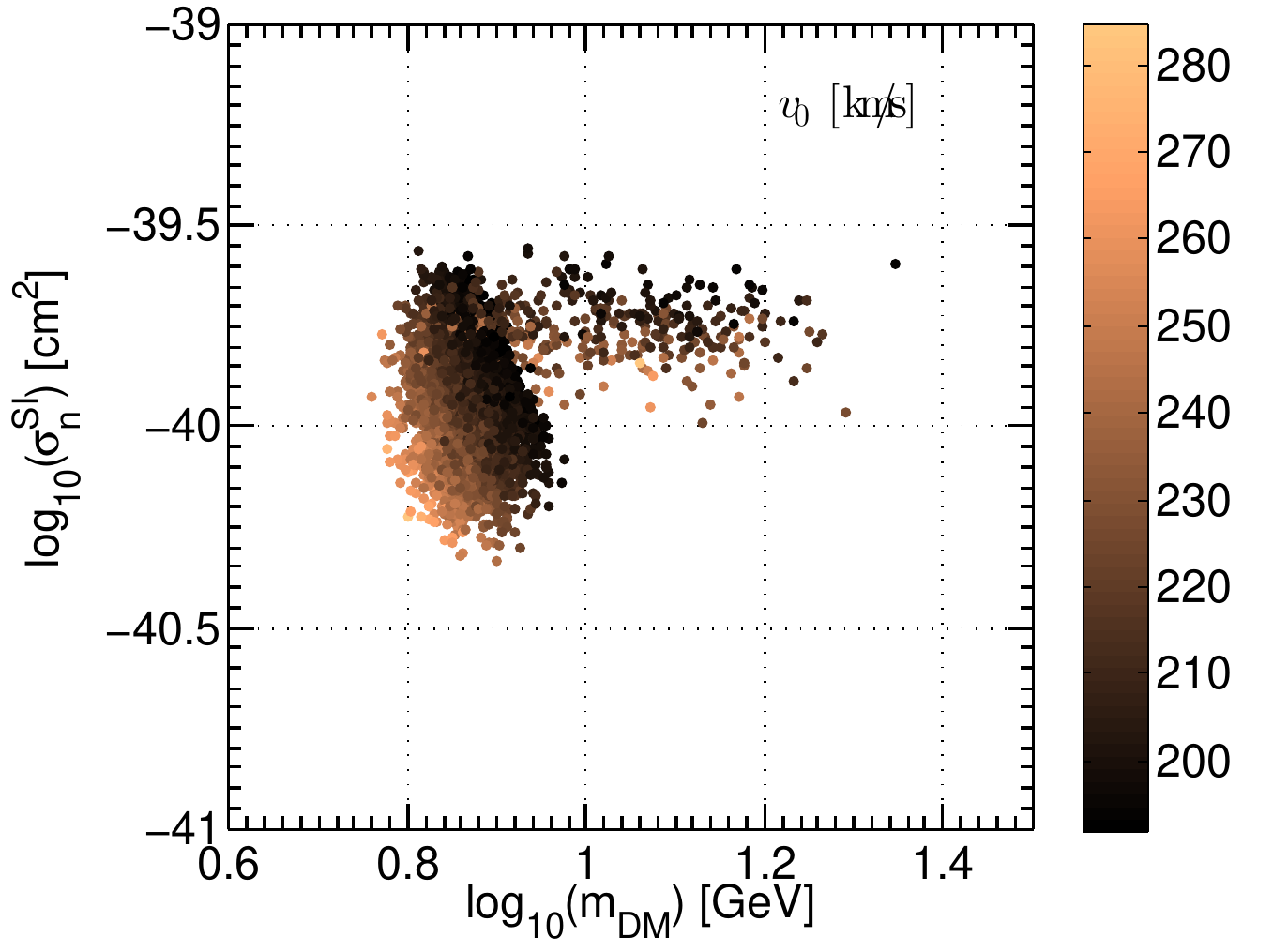}
\end{minipage}
\caption{{\it Left:} Results for a combined fit of the detection claims. 2D marginal posterior in the \{$m_{\rm DM},\sigma_n^{\rm SI}$\}-plane for the individual experimental regions and for the global fit combined in a single plot. The lines for DAMA (shaded), CoGeNT (cyan), CRESST (blue) and combined fit (red dashed) denote the 90\% and 99 \% credible regions, while for CDMS-Si only the 68\% is shown. {\it Center:} 1D pdf for the $q_{Na}$ nuisance parameter. The solid (dashed) lines stand for the posterior pdf (profile likelihood) and black (light blue) curves are for the combined  (DAMA alone) analysis. {\it Right:} 3D marginal posterior for DAMA and CoGeNT for \{$m_{\rm DM},\sigma_n^{\rm SI}\}$ and the circular velocity $v_0$. The third parameter direction is represented by the color code. The SI elastic scattering scenario and NFW density profile for the astrophysics (marginalized over) are assumed in all plots.}
\label{fig:combin}
\end{figure}
We have considered only the experiments that have hints of detection and analyzed them individually and combined together to find what is the best model that accounts for the data, among the SI scenarios proposed in Sec.~\ref{sec:dd}: elastic scattering (model 0, the simplest one with only $m_{\rm DM}$ and $\sigma_n^{\rm SI}$ as free parameters), inelastic scattering (model 1, standard free parameters plus $\delta >0$), isospin violating (model 2, standard free parameters plus $f_n/f_p$) and exothermic dark matter (model 3, standard free parameters plus $\delta <0$). In the right panel of Fig.~\ref{fig:lnbf} we show the outcome for model comparison against the simplest model. In DAMA, CRESST and CDMS-Si data, the support for the additional parameter is inconclusive as all the Bayes factors range from 0 to $|1|$: for instance a Bayes factor of 0.51 means that the isospin violating model is favored over the elastic scenario with the odds of 2:1 only. Conversely the CoGeNT data show moderate evidence against inelastic or exothermic dark matter, while the comparison is still inconclusive for isospin violating dark matter. The combined fit is driven by CoGeNT data: to explain simultaneously the hints of detection both SI elastic scattering or isospin violating dark matter are viable possibilities, on the contrary of inelastic and exothermic dark matter which are mildly disfavored with respect to the simplest model. The CoGeNT data are quite constraining and they do not support the extra parameter $\delta$, another example of Occams' razor at work, as the likelihood does not improve enough to compensate the volume increase due to $\delta$. A Bayes factor of -4.81 means that the inelastic scenario is disfavored with respect to the elastic case with the odds of 123:1, while the exothermic scenario is disfavored with the odds of 55:1. These scenarios as common explanation for the hints were already appearing disfavored in Fig.~\ref{fig:soa}.

We review an example of inference for the combined fit  and SI elastic scattering in Fig.~\ref{fig:combin} (red dashed contours, left panel). The combined region lies in between the credible regions of the single experiments. However, the best-fit point of the combined fit corresponds to a mass of $\sim 7$~GeV, compatible with CoGeNT but different from the best fit selected by DAMA at 15 GeV and the one of CDMS-Si, and a cross-section of $1.51 \times 10^{-40}\ {\rm cm}^2$, which is not a significant shift from the value selected by each experiment alone.  This fit comes also at the expense of a large shift in the circular velocity: $v_0=214^{+33}_{-1}  \ {\rm km \ s}^{-1}$ (90\% credible interval)  from the combined fit, versus  $v_0=229^{+36}_{-21}  \ {\rm km \ s}^{-1}$ from fitting for instance either DAMA, CoGeNT, CDMS-Si or CRESST alone, as shown in the right panel of Fig.~\ref{fig:combin} (DAMA and CoGeNT only). The preferred local dark matter density $\rho_\odot$ and escape velocity $v_{\rm esc}$ also suffer a downward shift, although not a significant one in either case. Commenting more on the DAMA sodium quenching factor $q_{\rm Na}$, it was previously an unconstrained quantity (light blue lines), while now it shows a preference for high values (black curves) right at the prior boundary $q_{\rm Na}=0.6$, see the central panel of Fig.~\ref{fig:combin}. This suggests that if a wider prior range for $q_{\rm Na}$ was allowed, an even higher value might have been preferred. This result is consistent with previous suggestions that a higher value for $q_{\rm Na}$ could improve the compatibility of DAMA and for instance CoGeNT data. To illustrate the impact of the quenching factor of DAMA on the combined fit we reduce the range of the quenching factor to $q_{\rm Na} = 0.2 \to 0.4$, which is more in line with the experimental measurements. The most recent measurement of the Na quenching factor indeed tends to prefer $q_{\rm Na} \sim 0.2$~\cite{Collar:2013gu}. The Bayesian procedure will always find a common region of compatibility, however this time the best fit point will have a much worst fit, as for instance the circular velocity shifts to the preferred value of $v_0 = 173^{+33}_{-1}  \ {\rm km \ s}^{-1}$. The sodium quenching factor will again show a peaked 1D marginalized posterior pdf and the location of the maximum will correspond to the highest allowed value for $q_{\rm Na}$. We refer the reader to~\cite{Arina:2011si} for details on this issue.

\begin{figure}[t!]
\begin{minipage}[t]{0.32\textwidth}
\centering
\includegraphics[width=1.\columnwidth,trim=10mm 15mm 13mm 14mm,clip]{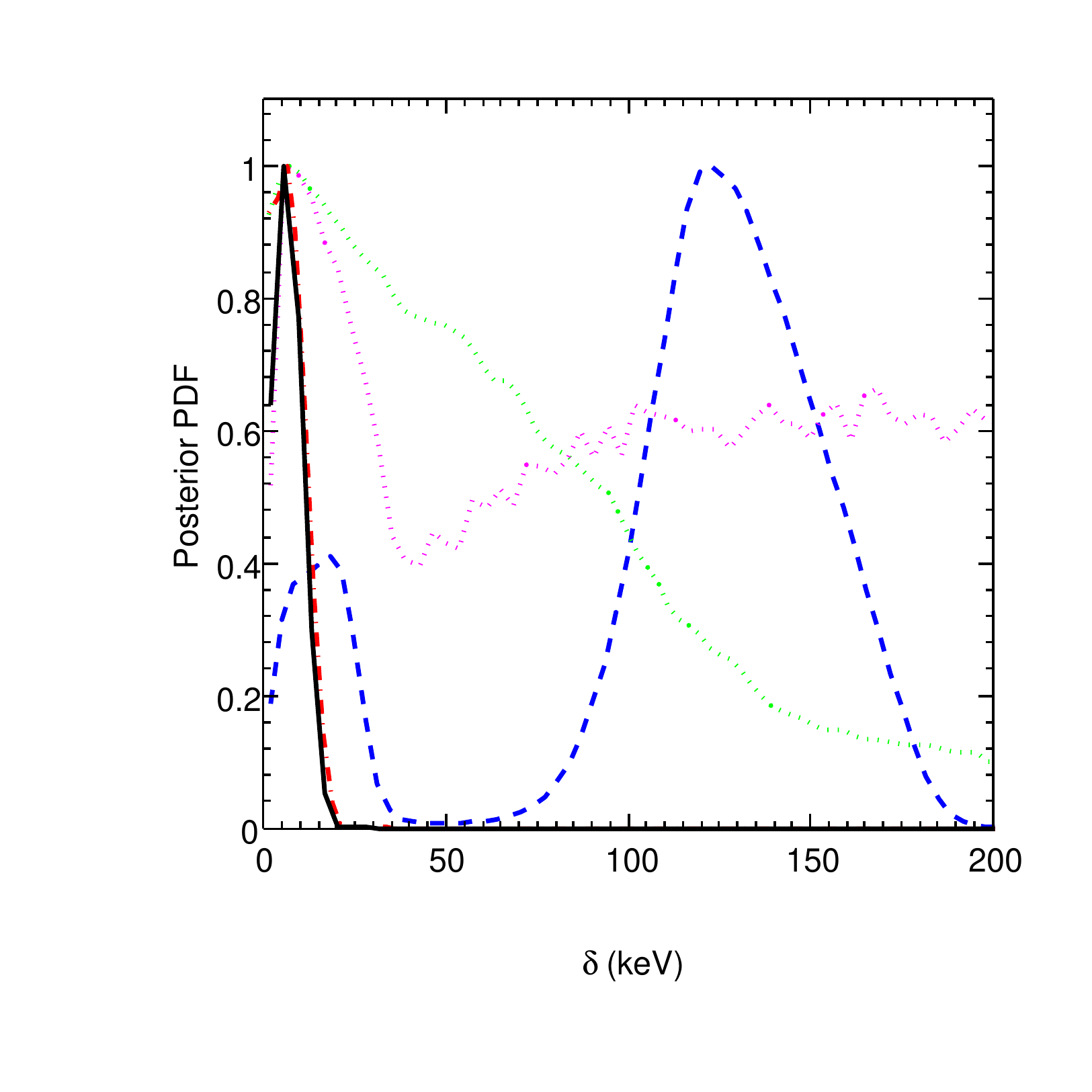}
\end{minipage}
\begin{minipage}[t]{0.32\textwidth}
\centering
\includegraphics[width=1.\columnwidth,trim=10mm 15mm 13mm 14mm,clip]{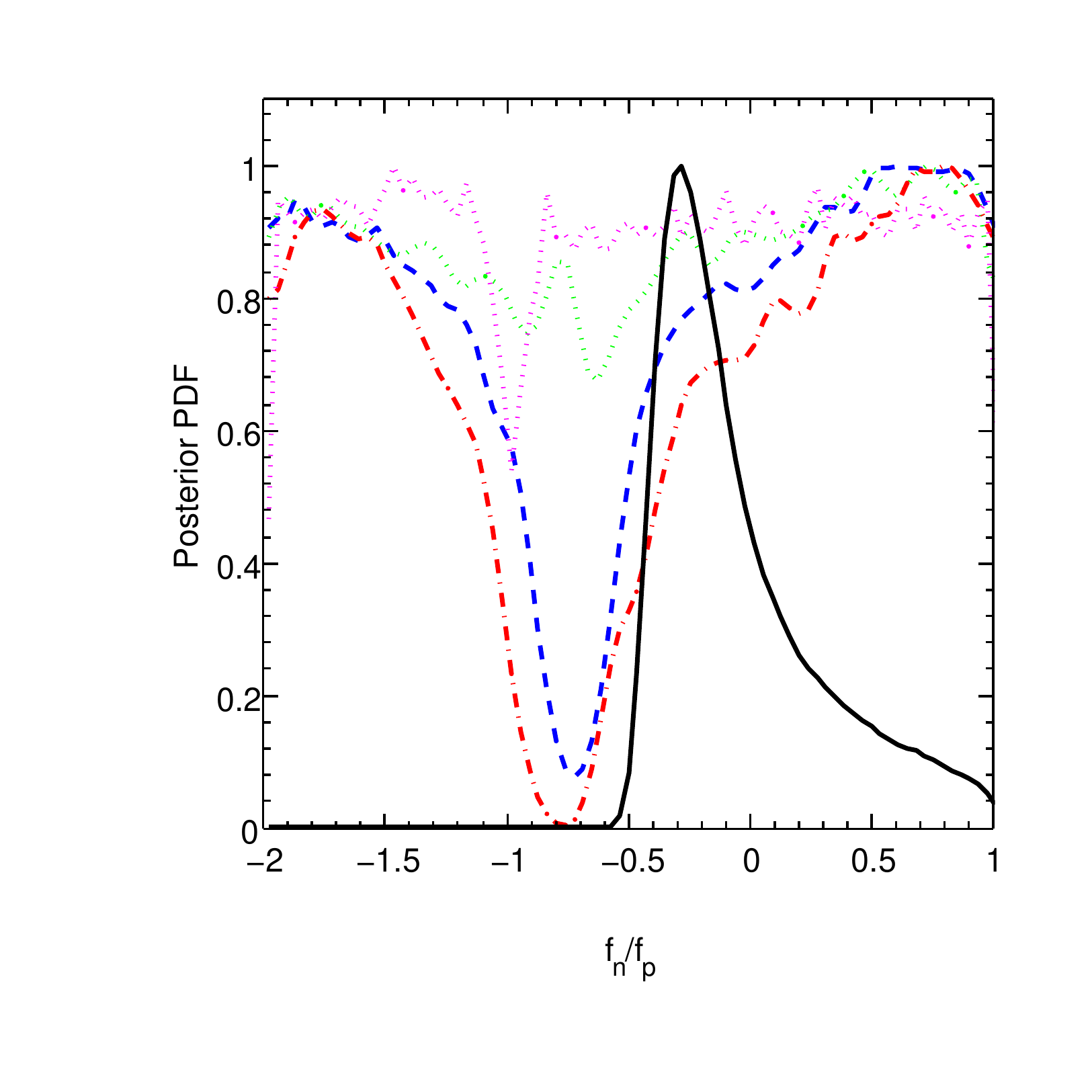}
\end{minipage}
\begin{minipage}[t]{0.32\textwidth}
\centering
\includegraphics[width=1.\columnwidth,trim=10mm 15mm 13mm 14mm,clip]{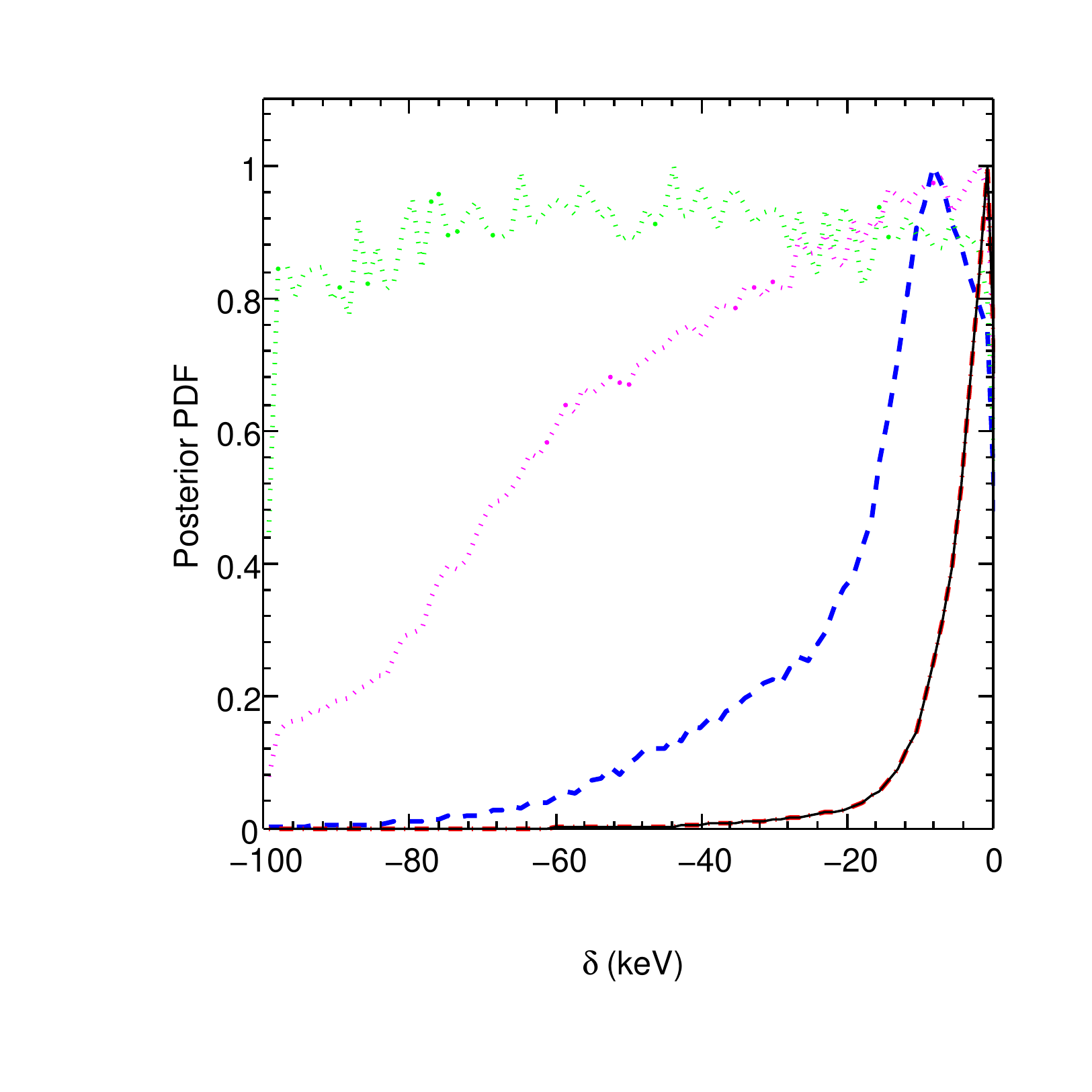}
\end{minipage}
\caption{{\it Left:} 1D marginal posterior pdf for the mass splitting $\delta$ in the inelastic dark matter scenario for DAMA (blue dashed), for CoGeNT (red dot-dashed), for CRESST (green dotted), for CDMS-Si (magenta dotted) and for the combined fit (black solid). {\it Center and right:} Same as left for the isospin violating parameter $f_n/f_p$ and for exothermic scattering $\delta$ respectively.}
\label{fig:1dpar}
\end{figure}

The nuisance parameters of the other experiments get affected by the combinations of all data and shift towards values that allow the best compromise but come at the detriment of having a good fit for the individual experiments. This is illustrated in Fig.~\ref{fig:1dpar}: the black line denotes the combined fit and shows that the mass splitting parameter $\delta$ (either positive panel, either negative right panel) is driven by CoGeNT and tends to be close to 0. This is a confirmation of the fact the CoGeNT data `like' elastic scattering, contrary to DAMA data. All the 1D posterior pdfs for the single experiments are consistent with the 2D marginal posterior pdfs shown in Fig.~\ref{fig:soa} and with the discussion underneath. For instance it is clear that the isospin violating scenario depends on the type of nucleus: the behavior of the 1D posterior pdfs for $f_n/f_p$ has a deep at the value in which the interaction with the element is suppressed. CDMS-Si (magenta dashed) and CRESST (green dashed) demonstrate once again their low constraining power: the 1D posterior pdfs are almost flat in all the prior ranges used for $f_n/f_p$ and exothermic scenarios. Interestingly CDMS-Si disfavors the inelastic SI scenario.

\subsection{Sensitivity analysis}\label{sec:sa}
In any good Bayesian analysis it is important to assess the robustness of the results with respect to reasonable changes in the prior choices. The phenomenological approach used usually in dark matter direct detection, while perfectly convenient for parameter inference, poses a problem for model comparison:  without the guidance of specific predictions, the odds for a more complex model can be made arbitrarily small by increasing the width of the priors on the additional parameters. 

By means of the SDDR, we first assess whether $\ln B$ for the models $2a$ and $2b$ have been artificially suppressed or not.  Consider for concreteness model $2a$. If we were to reduce the prior ranges for all three fractional modulation amplitudes to $S_{\rm m}^i = 0 \to 0.5$ from the default choice of $S_{\rm m}^i=0 \to 1$, then it follows from the SDDR definition that the Bayes factor  $\ln B$ in favor of model $2a$ would increase by approximately $\ln 2^3 \simeq 2.1$~units, bringing it to $\sim -1.06$ relative to model 0. In such a case, the model comparison between model $2a$ and the consistent DM model $1b$ would produce a weak evidence in favor of the latter scenario, while a comparison between models $2a$ and $1a$ would still favor moderately the latter.  Indeed, even with the prior ranges reduced further to  $S_{\rm m}^i = 0 \to 0.2$, model $2a$ would hardly overcome model $1b$. We can conclude that the general statement that dark matter models are preferred over other physics models is robust from a Bayesian point of view.

As far as it concerns the particle physics scenarios, we exemplify here the procedure by considering the inelastic scenario. The SDDR implies that a rescaling of a factor of 2 in the prior range for $\delta$ will affect the Bayes factor by $\ln 2 = 0.69$~units, changing a moderate to a weak evidence against this model for the combined fit, while {\it e.g.} for DAMA the Bayes factor still remains inconclusive. We conclude that by choosing physically motived priors the elastic model is the best motivated model together with the isospin violating scenario and this conclusion is robust against variation of prior range~\cite{Arina:2012dr}.

On the same vein, we note that  the astrophysics does not play a crucial role in model selection and its contribution in the Bayes factor cancels out, because the values for $\ln B$ for fixed or varying  astrophysics have similar values.

\subsection{Connection with classical hypothesis testing: from $\ln B$ to $\Delta \chi^2_{\rm eff}$}\label{sec:cht}
Classical hypothesis testing attempts to rule out the null hypothesis $\mathcal{H}_0$ by quantifying the probability of observing data as extreme or more extreme than what has been obtained. To this end the p-value $\wp$ is used:  small values of $\wp$ denote that the observed data are very improbable under the null. We stress that p-values are not probabilities for hypotheses but they are probabilities of obtaining more extreme data than observed assuming the null hypothesis is correct. In order to obtain the probability for a hypothesis one needs to take a Bayesian approach, as we have illustrated in this review with a couple of examples. The mapping of the test statistic onto a p-value requires in general a Monte Carlo simulation; analytic solutions exist only in special cases, and apply only under certain conditions. A popular choice for the test statistic  in the context of nested models is the profile likelihood ratio. If the likelihood in the $N$ additional parameters of the more complex model is Gaussian and unbounded, then Wilks' theorem~\cite{Wilks:1938} applies, meaning that the test statistic $\Delta \chi^2_{\rm eff}$ is asymptotically distributed as a $\chi^2$ with $N$ degrees of freedom. One of the conditions that validate the application of  Wilks' theorem is that the likelihood must be unbounded, namely the additional parameters of the more complex model cannot sit on the boundary of its parameter space. This is precisely the situation we encounter in both examples illustrated above: the modulation amplitudes $S_{\rm m}^i$ or $\delta$ are equal to zero under the null hypothesis. Hence, Wilks' theorem cannot be applied to obtain the p-value from the $\Delta \chi^2_{\rm eff}$, a procedure that is often pursued outside its domain of validity. For the case where
\begin{enumerate}
\item[1.] the $N$ additional parameters are bounded, with the null hypothesis sitting on the boundary, 
\item[2.] the likelihood is Gaussian,
\item[3.] all parameters are identifiable under the null hypothesis, 
 \end{enumerate}
 Chernoff's theorem~\cite{Chernoff:1954,Shapiro:1988} is valid instead to evaluate the p-value of the null, and states that the distribution of the test statistic $\Delta \chi^2_{\rm eff}$ under the null is asymptotically a weighted sum of random variables $\chi^2_i$ following chi-squared distributions with $i$ degrees of freedom.

As for the annual modulation in CoGeNT,  Chernoff's theorem applies to compute the p-value of the null hypothesis of no modulation when the more complex hypothesis is identified with the dark matter  models $1a$ or $1b$. We obtain a p-value of $0.02$ under the null for all energy bins combined, when the alternative hypothesis is model $1a$. This corresponds to $\Delta\chi^2_{\rm eff} = 6.26$ and a $2.3\sigma$ detection\footnote{Assuming a Gaussian distribution to convert p-values into the number of sigmas.}. If instead we take model $1b$ as the alternative, the p-value is $0.1$, equivalent to a $1.6\sigma$ detection (and $\Delta \chi^2_{\rm eff} = 3.84$). However Chernoff's theorem cannot be applied to the other physics models $2a$ and $2b$, because these models contain parameters that are undefined  under the null: when $S_{\rm m}^i = 0$  the parameters $t_{\rm max}$ and $T$ are meaningless. Monte Carlo simulations would be required to determine the distribution of the test statistic when the alternative model is either $2a$ or $2b$ but for which $\Delta\chi^2_{\rm eff}$ is $10.63$ and $10.83$ respectively. 

Chernoff's theorem holds as well in the case of model comparison of the particle physics scenarios: the hypothesis of inelastic scattering as explanation for all hints of detection leads to $\wp = 0.001$, hence is rejected at $3.2\sigma$ CL ($\Delta \chi^2_{\rm eff} = 9.7$ for the additional new parameter), while exothermic dark matter is found to be disfavored at  $3.1\sigma$ CL  ($\Delta \chi^2_{\rm eff} = 8.1$) corresponding to $\wp = 0.002$. Details for direct detection are provided in~\cite{Arina:2011zh,Arina:2012dr}, while for statistics we refer to~\cite{Protassov:2002sz}. 

\section{Compatibility with XENON100 exclusion bound}\label{sec:comp}

First, as a new little exercise for this review, we assess what is the compatibility with the exclusion bound of XENON100 and the last claim of detection by CDMS-Si, by taking the ratio of the evidence of the combined fit (XENON100+CDMS-Si, with the hypothesis that they are compatible) divided by the product of the evidence for each experiment alone, Eq.~(\ref{eq:rtest}). The outcome of this ratio is $\mathcal{R}=-0.8$, which is inconclusive to either support or reject the hypothesis of compatibility between the two experiments. This is compatible with the Bayesian inference for CDMS-Si, where already the 90\% credible intervals does not delimit a closed region. 
\begin{figure}[t!]
\centering
\includegraphics[width=0.5\columnwidth,trim=25mm 63mm 25mm 65mm,clip]{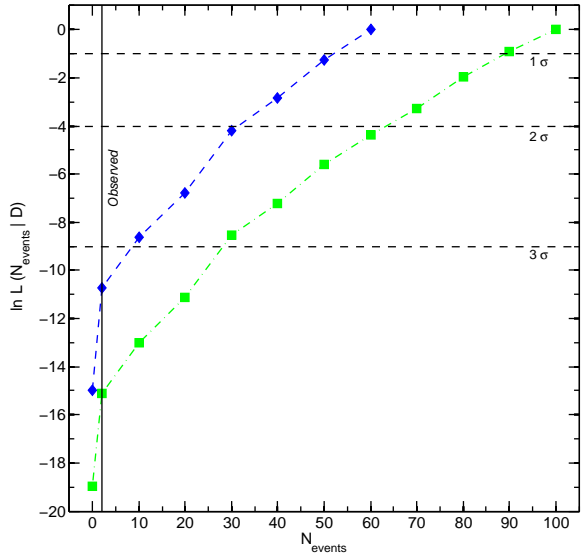}
\caption{Predictive data distribution ($\mathcal{L}-$test) for the number of events $N_{\rm events}$ in XENON100 detector. The curve represents the conditional evidence of XENON100 and the combined set ($\mathcal{D}=\{{\rm DAMA, CoGeNT, CRESST}\}$) at a given data point, divided by the maximum of the probability, for elastic SI interaction. The blue dashed curve is for a maximum of 60 events, while the green dot-dashed line stands for $N_{\rm max}=100$. The vertical line gives the actual measured value $N_{\rm obs}=2$. The data points denote the location at which the predictive probability has been computed and the lines are spline interpolation between those points. The horizontal dashed lines represent the $1,2$ and $3 \sigma$ significance. }
\label{fig:tests}
\end{figure}
Likewise, if the hypothesis $\mathcal{H}_0$ is the compatibility between XENON100 and the hints of detection combined together ($\mathcal{D}=\{ \rm DAMA, CoGeNT, CRESST\}$) discussed in~\cite{Arina:2012dr}, the $\mathcal{R}$-test, shows an inconclusive evidence against it; this means that the data are not constraining enough for a decisive outcome.

The $\mathcal{L}-$test, Eq.~(\ref{eq:ltest}), is found to have a more stringent outcome, being a likelihood ratio. The variable characterizing this test is the number of observed events in the XENON100 experiment, $\mathcal{T} \equiv N_{\rm events}$. To exemplify how it works, we have evaluated the conditional evidence $p(\mathcal{T}|\mathcal{D})$ and computed the predictive probability on a grid of values for $N_{\rm events}$. The relevant quantity $\mathcal{\ln L}(N_{\rm events}|\mathcal{D})$ is plotted in Fig.~\ref{fig:tests} as a function of the possible outcome of the experimental observation, with the actual observed value denoted by the solid black vertical line. Consider first the blue line/diamonds: the predictive probability grows fast increasing the number of events seen in the detector. This indicates that actually the compatibility of this experiment with the data $\mathcal{D}$ increases augmenting the number of events seen in XENON100. In other words, more than 2 events should have been observed for $\mathcal{T}$ and $\mathcal{D}$ to be consistent. We have found that the maximum of the probability depends on the number of events we assume have been seen, as the likelihood function is an increasing function of the observed number of events. If we consider $N_{\rm events}$ observed, the number of observed events that maximizes the likelihood function is $N_{\rm max} = N_{\rm events}$. Considering $N_{\rm max}=60$ the discrepancy between the data sets $\mathcal{D}$ and $\mathcal{T}$ is larger than $3\sigma$. Augmenting the number of observed events in the detector (green line and square) with $N_{\rm events}=100$ would lead to discrepancy larger than $4\sigma$. The predictive probability for the inelastic SI scattering scenario has the opposite behavior as the finest agreement between XENON100 and the combined fit is found for 0 observed events, result supported by the parameter inference shown in Fig.~\ref{fig:soa}. The isospin violating and exothermic scenarios follow closely the behavior of elastic scattering, although the discrepancy is marginal, at the level of $2\sigma$ for $N_{\rm max}=60$.  The outcome of this test is compatible with the analysis of the XENON100 collaboration~\cite{Aprile:2013teh}. The outcome of the likelihood ratio in data space means that the experimental results are incompatible (or marginally incompatible) under the hypothesis of dark matter and either the theoretical hypothesis should be changed or a closer look to systematics/uncertainties is appropriate.

\section{Conclusions and Perspectives}
\label{sec:concl}

Bayesian statistics offers a consistent framework to deal with uncertainties in several different situations, from parameter inference to model comparison, and can be applied to  generic observations,  both repeatable and one-off experiments. In this review we have illustrated its application to dark matter direct searches, which are affected by astrophysical uncertainties and experimental systematics.

We briefly resume our findings. Starting with parameter inference, in the simplest particle physics scenario, which is SI elastic scattering, the Bayesian results are compatible with classical statistical methods (see {\it e.g.}~\cite{Kelso:2011gd,Fornengo:2011sz,Kopp:2011yr,Frandsen:2011ts,Frandsen:2013cna}). The inclusion of nuisance parameters (with MCMC or nested sampling methods) and their consecutive marginalization does not alleviate the tension between hints of detection and the $90\%$ CL of the exclusion bound. We note that inference for CDMS-Si leads to a different result with respect to the profile likelihood analysis: the Bayesian 90\% credible region does not close, weakening considerably the claimed tension with XENON100. Considering the alternative particle physics scenarios, as inelastic, isospin violating or exothermic scattering, we have provided the full 2D posterior pdf marginalized over the additional theoretical parameter. This new parameter is not supported yet by the direct detection data, hence from inference point of view, it gives a relevant volume effect to the 2D marginalized posterior pdf of each experiment, augmenting the degree of compatibility between experimental results. Particularly relevant is the case of isospin violating dark matter.

Bayesian model selection, which is not possible with classical statistical methods, gives a quantitative answer on which is the best model that accounts for the data. Regarding the optimal model that accounts for the hints of detection, the Bayesian evidence has shown to moderately disfavor both inelastic and exothermic dark matter, while the outcome is inconclusive among elastic and isospin violating scenario. Concerning the evidence for an annual modulation signal in CoGeNT data to be due to WIMPs, we find that there is weak evidence for a modulation. Modulation models due to other physics, which vary the phase and period of the time-dependent signal, compare unfavorably with the no modulation case, paying the price for their excessive complexity. These model comparison conclusions are borne out both from a Bayesian model comparison point of view and from a classical hypothesis testing perspective. The modulation has been confirmed by the new scientific run of CoGeNT~\cite{Aalseth:2014nda}, even though with a smaller statistical significance with respect to the previous run, by means of a profile likelihood analysis described in details in~\cite{Aalseth:2014jpa}.

It is crucial to use appropriate statistical tools to assess the degree of compatibility of data sets in direct searches, because the significance of these observations is in the potential `discovery' or `rejection' zone of $\sim 3\sigma$. Testing the hypothesis of compatibility of data (XENON100 and the detection data sets) the outcome of the Bayesian analysis is inconclusive, while it would have been rejected by classical p-values. For instance within the framework of isospin violating dark matter the Bayes factor of XENON100 combined with the hints with respect to the single experiment is inconclusive, while the hypothesis of compatibility is discarded at $2\sigma$ CL by the likelihood ratio method. This is an example of Lindley's paradox, when Bayesian model selection returns a different result from classical hypothesis testing, see~\cite{Trotta:2005ar}. In general is has been shown that in presence of poorly constraining data, Bayesian statistics gives more robust constraints, see {\it e.g.}~\cite{loredo}. This simply implies that we will have to wait for more data and detectors with increased level of sensitivity to settle the debate (if the debate can be settled at all). In this respect there is a huge effort deployed from the experimental side: DAMA is running in its new configuration (called DAMA/LIBRA-phase2) with a lower software energy threshold aiming to improve the experimental sensitivity~\cite{Bernabei:2013xsa}. This low threshold at 2 keVee might shed light on the origin of the modulated signal as the data are expected to have a higher constraining power. Being more constraining they will also decrease the compatibility between experiments, as pointed out in~\cite{Kelso:2013gda}. CDMSlite~\cite{Agnese:2013lua}, which is a particular setup of the SuperCDMS experiment dedicated to low mass WIMP searches, has already released the analysis of the first preliminary run, without finding an excess over the background. The ton-scale detector XENON1T is already in construction and it is expected to probe the WIMP parameter space down to $\sigma \sim 10^{-47} \rm cm^2$.  For a review on the experimental future of direct searches we refer to~\cite{Baudis:2012ig}. We stress that if the experimental collaborations would in the future provide public likelihood codes, this would have a positive impact for the phenomenological analysis.

Concluding with an optimistic view in case of a consolidated dark matter detection, one might want to know how well the theoretical parameters, namely the WIMP mass and cross-section, can be inferred from the data. The reconstruction of such parameters can be improved for instance using multiple targets~\cite{Pato:2010zk,Cerdeno:2012ix}, however there are intrinsic limitations in direct searches due to $1/m_{\rm DM}$ dependence of the rate~\cite{Green:2007rb,Kavanagh:2012nr,Strege:2012kv}, that deteriorate the sensitivity to WIMP masses larger than $\sim 150$ GeV. The search for WIMPs has the advantage of having other identification strategies, such as indirect detection or collider searches, which are all highly complementary to direct detection, see {\it e.g.}~\cite{Arina:2010wv,Arina:2010an,Bertone:2010rv,Galli:2011rz,Arina:2013jya} and break degeneracies in the parameter space.  For a quite accurate reconstruction of the WIMP properties, it will be possible to improve the understanding of the astrophysics, {\it e.g.} reconstruct the underlying dark matter velocity distribution shape~\cite{Pato:2012fw}.

\section*{Acknowledgements}
The author warmly thanks F. Kahlhoefer for useful discussion on the CDMS-Si data, R. Trotta for discussion on the LUX likelihood, and acknowledges the support of the European Research Council through the ERC starting grant {\it WIMPs Kairos}, PI G. Bertone, and the partial support of the ERC project 267117 (DARK) hosted by Universit\'e Pierre et Marie Curie - Paris 6, PI J. Silk.

\appendix
\section{Likelihood for CDMS-II silicon data}\label{sec:app}
The cryogenic CDMS experiment at the Soudan Underground Laboratory operated germanium and silicon solid-state detectors.  Three events were 
observed at 8.2, 9.5 and 12.3 keVnr in the silicon run comprising 8 detectors made by $0.6$~kg, for a total exposure of 140.2 kg-days prior application of WIMP selection criteria~\cite{Agnese:2013rvf} (scientific run from July 2007 to September 2008). The expected background is $B_n = 0.13$  neutrons, $B_e=0.41$  electrons and $B_{Pb} = 0.08$ lead recoils from $^{210}$Po decay in the $7\to100$~keVnr detection window. The collaboration pursued a maximum likelihood analysis, finding that the probability that the known backgrounds would produce three or more events in the signal region is 5.4\%.

In order to exploit all available informations we use as well the spectral information. The likelihood function is then a product of three Poisson distributions stating the probability of seeing one event at $E_j$ ($j=1,2,3$ denotes the energy of each event) and a series
of Poisson distributions stating the probability of seeing zero events, for those energies with no events~\cite{loredo}, that is
\begin{equation}
\label{eq:lcdmsge}
\ln{\cal L}_{\rm CDMS-Si}  = -S-B + \sum_{j=1,3}  \ln \left( \frac{\der R}{\der E_j} + \frac{B_i}{\bar{B_i}}\frac{\der N_{B_i}}{\der E_j} \right)\,,
\end{equation}
where $S$ is the total expected signal in the recoil energy window and the index $i$ runs over the three sources of background, $i=n,e,Pb$ (the sum over $i$ is implicit). $B$ is the total background expected in the same energy window, the best fit value $\bar{B}=0.7$ events. However the pdf for the electron background, coming from surface events, follows a skewed gaussian distribution with peak value at $B_e=0.41$ and width given by $\sigma_e =  (-0.08+0.20)\,  {\rm stat}  + (-.24 + 0.28)\,  {\rm syst}$. To account for this distribution, we define  $N_e$ as a free normalization for the electron background, Gaussian distributed with mean $B_e$ and deviation $\sigma_e$. This is the only nuisance parameters that characterize the CDMS-Si likelihood, as for instance no quenching factor is required for the CDMS experiment, {\it i.e.} $q=1$. The spectral distribution $\der N_{B_i}/\der E$ of each background can be found in~\cite{McCarthy:2013gya}. The surface electron background is exponentially falling off in the energy window for WIMP detection, while the neutron and lead background are constant in the same energy range.

\bibliographystyle{elsarticle-num.bst}
\bibliography{biblio}

\end{document}